\documentclass{llncs}
\usepackage{amsmath}

\usepackage{amsmath}
\usepackage{amssymb}
\usepackage{multicol}
\usepackage{verbatim}
\usepackage{enumitem}
\usepackage{graphicx}
\usepackage{algorithm}
\usepackage{algpseudocode}
\usepackage{mathtools}
\usepackage{multirow}
\usepackage{bm}
\usepackage{nicefrac}
\usepackage{xspace}
\usepackage{moresize}

\usepackage{float}
\floatname{algorithm}{\small Algorithm}
\newfloat{algorithm}{t}{lop}

\input{macros.tex}

\newcommand{\sss}{\hspace{0.1in}}
\newcommand{\texttts}[1]{\texttt{\small #1}}

\makeatletter
\newcommand*{\da@rightarrow}{\mathchar"0\hexnumber@\symAMSa 4B }
\newcommand*{\da@leftarrow}{\mathchar"0\hexnumber@\symAMSa 4C }
\newcommand*{\xdashrightarrow}[2][]{%
  \mathrel{%
    \mathpalette{\da@xarrow{#1}{#2}{}\da@rightarrow{\,}{}}{}%
  }%
}
\newcommand{\xdashleftarrow}[2][]{%
  \mathrel{%
    \mathpalette{\da@xarrow{#1}{#2}\da@leftarrow{}{}{\,}}{}%
  }%
}
\newcommand*{\da@xarrow}[7]{%
  \sbox0{$\ifx#7\scriptstyle\scriptscriptstyle\else\scriptstyle\fi#5#1#6\m@th$}%
  \sbox2{$\ifx#7\scriptstyle\scriptscriptstyle\else\scriptstyle\fi#5#2#6\m@th$}%
  \sbox4{$#7\dabar@\m@th$}%
  \dimen@=\wd0 %
  \ifdim\wd2 >\dimen@
    \dimen@=\wd2 %
  \fi
  \count@=2 %
  \def\da@bars{\dabar@\dabar@}%
  \@whiledim\count@\wd4<\dimen@\do{%
    \advance\count@\@ne
    \expandafter\def\expandafter\da@bars\expandafter{%
      \da@bars
      \dabar@ 
    }%
  }%
  \mathrel{#3}%
  \mathrel{%
    \mathop{\da@bars}\limits
    \ifx\\#1\\%
    \else
      _{\copy0}%
    \fi
    \ifx\\#2\\%
    \else
      ^{\copy2}%
    \fi
  }%
  \mathrel{#4}%
}
\makeatother

\newcommand{\toolname}{{\sc Optix}\xspace}

\begin{document}
\mainmatter              
\title{Eventually Sound Points-To Analysis with Missing Code}
\titlerunning{Eventually Sound}  
%
\author{Osbert Bastani\inst{1} \and Lazaro Clapp\inst{1} \and Saswat Anand\inst{1} \and Rahul Sharma\inst{2} \and Alex Aiken\inst{1}}

\authorrunning{Osbert Bastani et al.} 


\institute{Stanford University, USA,\\
\email{obastani@cs.stanford.edu, lclapp@stanford.edu, saswat@cs.stanford.edu, aiken@cs.stanford.edu}
\and
Microsoft Research, India,\\
\email{rahsha@microsoft.com}}

\maketitle              

\begin{abstract}
Static analyses make the increasingly tenuous assumption that all source code is available for analysis; for example, large libraries often call into native code that cannot be analyzed. We propose a points-to analysis that initially makes optimistic assumptions about missing code, and then inserts runtime checks that report counterexamples to these assumptions that occur during execution. Our approach guarantees \emph{eventual} soundness, i.e., the static analysis is sound for the available code after some finite number of counterexamples. We implement \toolname, an eventually sound points-to analysis for Android apps, where the Android framework is missing. We show that the runtime checks added by \toolname incur low overhead on real programs, and demonstrate how \toolname improves a client information flow analysis for detecting Android malware.
\end{abstract}

\section{Introduction}
\label{sec:introduction}

To guarantee soundness, static analyses often assume that all program source code is available for analysis. This assumption has become tenuous as programs increasingly depend on large libraries and frameworks that are prohibitively difficult to analyze. For example, mobile app stores can use static analysis to improve the quality of published apps by searching for malicious behaviors~\cite{fuchs2009scandroid,feng2014apposcopy,arzt2014flowdroid} or security vulnerabilities~\cite{fahl2012eve,mutchler2016target,egele2013empirical}. However, Android apps depend extensively on the Android framework, which makes frequent use of native code and reflection, both of which are practical barriers to static analysis. Therefore, the Android framework is often omitted from the static analysis~\cite{zhu2013automated,bastani2015specification}, in which case we refer to it as \emph{missing}. In any large software system, there are inevitably parts that are missing and cannot be handled soundly~\cite{livshits2015defense}.

When code is missing, one of the following desirable properties must be sacrificed: (i) soundness (by making optimistic assumptions), (ii) precision (by making pessimistic assumptions), or (iii) automation (by using human-written \emph{specifications} that summarize missing code)~\cite{zhu2013automated,bastani2015specification}. For many analyses, pessimistic assumptions are simply too imprecise, and losing soundness is a significant compromise. For example, consider malware detection---a security analyst must examine every potential malware, making false positives costly, but unsoundness can be exploited by a knowledgeable attacker to avoid detection.

Using specifications is a promising compromise---in principle, for a one-time cost of writing specifications, the precision of the analysis can be greatly improved without sacrificing soundness. However, specifications are costly to write, and furthermore must be updated whenever the missing code changes. Oftentimes, an effective strategy is to implement specifications as needed---in the malware example, large portions of libraries are typically irrelevant to the static analysis. Of course, determining which specifications are relevant can be very error prone. Typically, missing specifications are optimistically assumed to be empty, leaving open the possibility of false negatives. These tradeoffs can be alleviated, but not eliminated, by \emph{inferring} specifications, e.g., automatically based on dynamic information~\cite{clapp2015modelgen,heule2015mimic} or interactively with a human analyst~\cite{zhu2013automated,bastani2015specification}.

We propose a novel appraoch that may offer better tradeoffs. Given a program (e.g., an Android app), we first run the static analysis with optimistic assumptions about missing specifications. If no errors are found, then the program is instrumented to detect counterexamples to the optimistic assumptions, and the instrumented program is published (e.g., on Google Play). If a counterexample is ever detected, then it is reported back to the publisher (e.g., Google), who can update their specifications and re-run the static analysis; at this point, the program may also be terminated (e.g., to ensure that no malicious functionality is executed). With an appropriate instrumentation scheme, our approach satisfies three important properties:
\begin{itemize}
\item {\bf Eventual soundness}: If any counterexample occurs during execution, then the program instrumentation reports some counterexample. Furthermore, only finitely many such reports will ever be issued.
\item {\bf Precision:} The analysis is at least as precise as having all specifications available.
\item {\bf Automation:} The analysis is highly automated.
\end{itemize}
The key property of interest is eventual soundness. The first part of this property is analogous to the benefit of dynamic type systems or dynamic information flow control~\cite{austin2012multiple,de2012flowfox,enck2014taintdroid}---issues are caught as soon as they occur, thereby minimizing potential damage (at the cost of some runtime overhead). The second part says that eventually, the static analysis becomes sound, at least with respect to all remaining executions. Existing program analyses either have a formal soundness guarantee or are unsound analyses with no formal results. Eventually sound analyses are potentially unsound but are only a finite number of counterexamples away from achieving provable correctness.    

%
%

To design an eventually sound program analysis, we must design an instrumentation scheme that satisfies the eventual soundness property. Schemes satisfying eventual soundness or similar properties have been proposed for type checking~\cite{flanagan2006hybrid}, resolving reflective call targets~\cite{bodden2011taming}, and determining reachable code~\cite{bastani2015interactively}. However, in these settings, the schemes are relatively straightforward---e.g., to detect missing reflective call targets, it suffices to record the call target of each reflective call.

In this paper, we propose an eventually sound points-to analysis for Android apps, where the Android framework is missing. We focus on points-to analysis since it lies at the core of many static analyses, and we believe that eventually sound clients (e.g., static information flow analysis) can be designed around our analysis. In our setting, counterexamples are \emph{missing points-to edges} that occur during an execution but are missing from the (optimistic) static analysis. Our main contribution is an eventually sound instrumentation scheme that detects and reports missing points-to edges. In contrast to previous settings, designing such a scheme for points-to analysis can be very challenging for two reasons:
\begin{itemize}
\item Na\"{i}vely using a dynamic points-to analysis to detecting counterexamples can incur huge overhead---for example, ~\cite{clapp2015modelgen} reports a 20$\times$ slowdown, and~\cite{mock2001dynamic} reports a slowdown of two orders of magnitude.
\item It is often not possible to insert runtime checks into missing code, e.g., in native code. Thus, we restrict our analysis to instrument only \emph{available} code.
\end{itemize}

To address the first challenge, we leverage the fact that to be eventually sound, we do not need to report \emph{every} counterexample that occurs during an execution. For example, if a \emph{potentially missing} points-to edge $x\hookrightarrow o$ can only occur during an execution if the potentially missing points-to edge $y\hookrightarrow o$ also occurs, then we only need to monitor whether $y\hookrightarrow o$ occurs. By leveraging this property, we substantially reduce the amount of required instrumentation. For programs where instrumentation in performance-critical parts is required, the overhead can be further reduced by sampling~\cite{liblit2003bug} or by manually adding specifications summarizing the relevant missing code.

For the second challenge, note that because we use specifications, we are already unable to discover relationships about the missing code. For many clients, only relationships between variables in the available code are of interest---e.g., Android malware can be characterized by relationships between variables in the app code alone~\cite{feng2014apposcopy}. However, these relationships typically depend on relationships between variables in the missing code. For points-to analysis, we cannot observe when variables in the app might be aliased because they both point to the same object allocated in missing code. To address this issue, our analysis introduces \emph{proxy objects} that correspond to concrete objects allocated in missing code,\footnote{The term \emph{proxy object} is ours, but the concept has occurred in prior work~\cite{bastani2015specification}.} which enable us to soundly and precisely compute client relations that refer only to available code (e.g., aliasing and concrete types).

We implement our eventually sound points-to analysis in a tool called \toolname\footnote{\toolname stands for Optimistic Points-To Information from eXecutions.}, which analyzes Android apps treating the entire Android framework as missing code. We show that our instrumentation typically incurs low overhead---the median overhead is 4.3\%, the overhead is less than 20\% for more than 90\% of the apps in our benchmark, and the highest is about 50\%. The overhead of the outliers can be reduced as described above; in particular, only a few manually provided specifications are needed to reduce the overhead of the outliers to reasonable levels (see Section~\ref{sec:overhead}). In summary, our contributions are:
\begin{itemize}
\item We propose a points-to analysis for programs with calls to missing code that is eventually sound, precise, and automatic (Section~\ref{sec:eventual}).
\item We minimize instrumentation to reduce runtime overhead (Section~\ref{sec:eventual}) and introduce proxy objects to handle allocations in missing code (Section~\ref{sec:proxy}).
\item We implement \toolname, a points-to analysis for Android apps that treats the entire Android framework as missing, and show that the instrumentation overhead is manageable (Section~\ref{sec:experiments}). The largest app that we have studied has over 300 thousand lines of Jimple code.
\end{itemize}

\section{Overview}
\label{sec:motivating}

\begin{figure}[t]
\begin{scriptsize}
\begin{minipage}{0.45\textwidth}
\begin{verbatim}
void main() { // program
  String str = mkStr();
  List list = new List(); // o_list
  list.set(str);
  Object data = list.get();
  if(randBool()) {
    Object dataCopy = data;
    sendHttp(dataCopy); }}
\end{verbatim}
\end{minipage}
\begin{minipage}{0.45\textwidth}
\begin{verbatim}
String mkStr() { // library
  String libStr = new String(); // o_str
  return libStr; }
void sendHttp(String str) { // library
  ... }
class List { // library
  Object f;
  void add(Object ob) { f = ob; }
  Object get(int i) { return f; } }
\end{verbatim}
\end{minipage}
\end{scriptsize}
\caption{Program \texttts{main} (left) calls various library functions, for which the analyst provides specifications (right). Abstract objects $o_{\text{list}}$ and $o_{\text{str}}$ are labeled in comments.}
\label{fig:main}
\end{figure}

Consider the program in Figure~\ref{fig:main}. Suppose that a security analyst asks whether the program leaks the return value of \texttts{mkStr} to the Internet via a call to \texttts{sendHttp}, which requires knowing that \texttts{str} and \texttts{dataCopy} may be aliased. We use points-to analysis to determine which variables may be aliased. In particular, a points-to analysis computes \emph{points-to edge} $x\hookrightarrow o$ if variable $x$ may point to a concrete object $\bar{o}$ allocated at allocation statement $o\in\mathcal{O}$ (called an \emph{abstract object}) during execution. Two variables may be aliased if they may point to the same abstract object. Our example program exhibits points-to edges such as $\texttts{list}\hookrightarrow o_{\text{list}}$, $\texttts{str}\hookrightarrow o_{\text{str}}$, and $\texttts{dataCopy}\hookrightarrow o_{\text{str}}$, so the points-to analysis concludes that $\texttts{str}$ and $\texttts{dataCopy}$ may be aliased.

Suppose that the library code is missing. For many clients (including static information flow analysis), it suffices to compute edges for \emph{visible} variables $x\in\mathcal{V}_P$ in the available code; however, these edges often depend on relationships in the missing code. Pessimistically assuming that missing code can be arbitrary is very imprecise, e.g., we may have $\texttts{data}\hookrightarrow o_{\text{list}}$ in case the implementation of $\texttts{get}$ is $\texttts{return this}$. Alternatively, optimistically assuming that missing code is empty can be unsound, for example, failing to compute $\texttts{data}\hookrightarrow o_{\text{str}}$ and $\texttts{dataCopy}\hookrightarrow o_{\text{str}}$. Such \emph{dynamic} points-to edges that are not computed statically are \emph{missing}.

A typical approach in practice is to provide \emph{specifications}, which are code fragments that overapproximate the points-to behaviors of library functions. Examples of specifications are shown in Figure~\ref{fig:main}. For instance, because our static points-to analysis collapses arrays into a single field, we can overapproximate array of elements stored by the \texttt{List} class as a single field \texttt{f}.

Suppose that the analyst has provided specifications for frequently used library functions such as \texttts{mkStr} and \texttts{sendHttp}, but a long tail of specifications remain missing, including those for \texttts{add} and \texttts{get}. Therefore, the (optimistic) static information flow analysis incorrectly concludes that \texttts{dataCopy} cannot point to \texttts{str}, and that \texttts{mkStr} therefore does not leak to the Internet. Furthermore, dynamic information flow control cannot be applied since the missing code cannot be instrumented without modifying every end user's Android installation.

Our analysis instruments the Android app to detect whether \emph{counterexamples} to the optimistic assumption that every missing specification is empty; this instrumentation only inserts runtime checks in the available code. The instrumented app is published on Google Play; if the instrumentation observes that a counterexample occurs during an execution, then it reports it back to the static analysis, which is recomputed to account for this new information.

Our example program \texttt{main} is instrumented to record the concrete objects pointed to by \texttts{libStr} and \texttts{data}. When the program is run:
\begin{itemize}
\item The variable \texttts{libStr} points to concrete object $\bar{o}_{\text{str}}$, so our analysis concludes that $\bar{o}_{\text{str}}$ is allocated at $o_{\text{str}}$.
\item The variable \texttts{data} points to $\bar{o}_{\text{str}}$, so our analysis concludes that $\texttts{data}\hookrightarrow o_{\text{str}}$ and reports this counterexample.
\end{itemize}
Upon receiving this report, we add $\texttts{data}\hookrightarrow o_{\text{str}}$ to the known counterexamples.

Given a new counterexample $x\hookrightarrow o$, the static analysis at the very least learns that $x\hookrightarrow o$ is a points-to edge that may occur. There are two ways in which the static analysis can generalize from this fact. First, it can compute additional missing points-to edges that are consequences of this fact according to the rules of the static analysis. For example, given the counterexample $\texttts{data}\hookrightarrow o_{\text{str}}$, our static analysis additionally computes its consequence $\texttts{dataCopy}\hookrightarrow o_{\text{str}}$, and determines that \texttts{str} and \texttts{dataCopy} may be aliased. Thus, the security analyst learns that the return value of \texttts{mkStr} may leak to the Internet, and can report any newly discovered bugs to the developer. In this case, the leak is discovered even if \texttts{randBool} returns false and the data is not leaked in that specific execution.

Second, the static analysis can also attempt to use \emph{specification inference} to try and identify which missing specification may have been the ``cause'' of the missing points-to edge. By doing so, the static analysis generalizes the counterexample to eliminate unsoundness when analyzing future apps. In Section~\ref{sec:infer}, we show how our tool leverages an existing specification inference algorithm to automatically infer candidate specifications that ``explain'' the counterexample. For example, given counterexample $\texttts{data}\hookrightarrow o_{\text{str}}$, the specification inference algorithm would infer the specifications for \texttt{add} and \texttt{get} shown in Figure~\ref{fig:main}. One caveat is that the inferred specifications must be validated by a human, since it is impossible to guarantee that they are correct. We show that in practice, the inference algorithm has high accuracy.

\begin{figure}[t]
\scriptsize
\begin{minipage}{0.45\textwidth}
\begin{enumerate}
\item (allocation) $\dfrac{\begin{array}{c}\\\\x\gets X(),~o=(x\gets X())\end{array}}{x\hookrightarrow o\in\Pi}$ \\
\item (assignment) $\dfrac{x\gets y,~y\hookrightarrow o\in\Pi}{x\hookrightarrow o\in\Pi}$ \\
\end{enumerate}
\end{minipage}
\begin{minipage}{0.5\textwidth}
\begin{enumerate}
\setcounter{enumi}{2}
\item (load/store) $\dfrac{\begin{array}{c}x\gets y.f,~z.f\gets w,~y\hookrightarrow o'\in\Pi,\\z\hookrightarrow o'\in\Pi,~w\hookrightarrow o\in\Pi\end{array}}{x\hookrightarrow o\in\Pi}$ \\
\item (missing) $\dfrac{x\hookrightarrow o\in\Pi_{\text{miss}}}{x\hookrightarrow o\in\Pi}$
\end{enumerate}
\end{minipage}
\caption{Rules to compute sound points-to sets. Rules 1-3 are standard. Rule 4 adds reported counterexamples to the analysis.}
\label{fig:pointsto}
\end{figure}

Next, we describe how our analysis instruments apps to detect missing points-to edges. Na\"{i}vely, we could use a dynamic points-to analysis, which instruments every allocation, assignment, load, and store operation in the program to determine all of the dynamic points-to edges that occur during an execution. However, this approach requires far more instrumentation than necessary. In particular, for eventual soundness to hold, the instrumentation must report only \emph{one} counterexample even if many counterexamples occur during an execution. If the analysis satisfies this property, because only finitely many counterexamples are reported, then after the last reported counterexample, the static analysis becomes sound for all subsequent executions. Leveraging this property enables us to substantially reduce the required instrumentation. For example, note that the missing points-to edge $\texttts{dataCopy}\hookrightarrow o_{\text{str}}$ can only occur during execution where $\texttts{data}\hookrightarrow o_{\text{str}}$ is reported, after which it is anyway computed by the static analysis. Therefore, we do not need to detect or report $\texttts{dataCopy}\hookrightarrow o_{\text{str}}$. 

Another challenge with the instrumentation is how to handle allocations in missing code. For example, if the specification for \texttts{mkStr} were also missing, then our analysis cannot instrument \texttts{libStr} to determine that $\bar{o}_{\text{str}}$ was allocated at $o_{\text{str}}$. Nevertheless, we can reason about such missing abstract objects based on observations in available code. In particular, suppose we instrument \texttts{str} and \texttts{list}. During execution, this instrumentation detects that \texttts{str} points to a concrete object $\bar{o}_{\text{str}}$. Since $\bar{o}_{\text{str}}$ was not allocated at $o_{\text{list}}$, it must have been allocated in \texttts{mkStr}. We represent this fact by introducing a \emph{proxy object} $p_{\text{mkStr}}$ pointed to by the return value $r_{\texttt{mkStr}}$ of \texttts{mkStr}. We discuss proxy objects in Section~\ref{sec:proxy}.

Finally, we describe an eventually sound points-to analysis, but more work is needed to ensure that the information flow client itself is eventually sound. We describe a candidate eventually sound information flow analysis in Section~\ref{sec:discussion}; evaluating this analysis is beyond the scope of our work.

\section{Eventually Sound Points-To Analysis}
\label{sec:eventual}

We describe our eventually sound points-to analysis, summarized in Figure~\ref{fig:algo}.

\subsection{Background and Assumptions}
\label{sec:assumptions}

Consider a program $P$ (whose code is available) containing calls to functions in a library $L$ (whose code is missing). There are five kinds of statements: allocations ($x\gets X()$, where $X\in\mathcal{C}$ is a class), assignments ($x\gets y$, where $x,y\in\mathcal{V}_P$ are program variables), loads ($x\gets y.f$, where $f\in\mathcal{F}$ is a field), stores ($x.f\gets y$), and calls to library functions $m\in\mathcal{M}$ library ($x\gets m(y)$). We omit control flow statements since our static analysis is flow-insensitive. We let $p_m$ (resp., $r_m$) denote the parameter (resp., return value) of library function $m$. For convenience, we assume that each library function has exactly one argument, and that there are no functions in $P$.

Our static points-to analysis, described in Figure~\ref{fig:pointsto}, is a standard flow- and context-insensitive analysis for computing points-to edges $\Pi\subseteq\mathcal{V}_P\times\mathcal{O}$~\cite{andersen1994program}; we describe how our results can be extended to context- and object-sensitive analyses with on-the-fly callgraph construction in Section~\ref{sec:contextsensitivity}. Rules 1-3 handle the semantics of each kind of statement. A function call $x\gets m(y)$ is treated as an assignment of $y$ to the parameter $p_m$ and an assignment of the return value $r_m$ to $x$. Rule 4 handles known counterexamples $\Pi_{\text{miss}}\subseteq\mathcal{V}_P\times\mathcal{O}$.

We initially make three simplifying assumptions. First, we assume that library functions do not contain allocations; we remove this assumption in Section~\ref{sec:proxy}. Second, we make the \emph{disjoint fields assumption}, which says that $\mathcal{F}_L\cap\mathcal{F}_P=\emptyset$, where $\mathcal{F}_L$ (resp., $\mathcal{F}_P$) are fields accessed by the library (resp., program), i.e., there are no \emph{shared fields} $f\in\mathcal{F}_L\cap\mathcal{F}_P$. We discuss how to weaken this assumption in Section~\ref{sec:sharedfields}. Third, the programs we consider do not have callbacks; we discuss how to handle callbacks in Section~\ref{sec:callbacks}.

\subsection{Eventual Soundness}

We first define soundness relative to an execution:
\begin{definition}
\rm
A points-to set $\Pi$ is \emph{sound} relative to an execution $e$ if no counterexamples occur, i.e., there is no dynamic points-to edge $x\hookrightarrow o\not\in\Pi$.
\end{definition}
Consider a points-to analysis that for a sequence of instrumented executions $E\triangleq e_1,e_2,\ldots$ computes a sequence of points-to sets $\Pi_E\triangleq\Pi_1,\Pi_2,\ldots$,  both indexed by the natural numbers $i\in\mathbb{N}$. Here, $\Pi_i$ is computed as a function of the previous points-to set $\Pi_{i-1}$ and counterexamples from $e_i$.
Note that the instrumentation for $e_{i+1}$ can be chosen adaptively based on $\Pi_i$ and that $\Pi_i\subseteq\Pi_j$ if 
$e_i,e_j$ is a subsequence of $E$.
\begin{definition}
\rm
The points-to analysis is \emph{eventually sound} if for any sequence $E$ of executions
$\Pi_i$ is sound relative to executions $e_1,\ldots,e_i$ and 
there exists a $N\in\mathbb{N}$ such that the sequence $\Pi_E$ has at most $N$ distinct  elements.
\end{definition}
A consequence of these definitions is that in an eventually sound points-to analysis only a finite number of executions can produce counterexamples. 
\begin{definition}
\label{defn:precise}
\rm
The points-to analysis is \emph{precise} if for every $i\in\mathbb{N}$, the points-to set $\Pi_i$ is a subset of the points-to set computed by analyzing the implementation of the missing code.
\end{definition}


Note that while progress towards soundness is guaranteed, it is not possible to report how many sources of unsoundness remain at any point in time; in general, even if all program paths are executed, there may be missing points-to edges.
For example, in the following code, suppose that \texttts{foo} is missing; then, if \texttts{randInt} never evaluates to \texttts{0}, the points-to edge $\texttts{y}\hookrightarrow o$ remains missing:


\noindent
\begin{center}
\begin{minipage}{0.45\textwidth}
\begin{scriptsize}
\begin{verbatim}
void main() { // program
  Object x = new Object(); // o
  Object y = foo(x); }
\end{verbatim}
\end{scriptsize}
\end{minipage}
\begin{minipage}{0.45\textwidth}
\begin{scriptsize}
\begin{verbatim}
Object foo(Object ob) { // library
  Object[] arr = new Object[2];
  arr[0] = ob;
  return arr[randInt()]; }
\end{verbatim}
\end{scriptsize}
\end{minipage}
\end{center}


Despite the inability to quantify progress, the property of eventual soundness is useful, since it guarantees that only a finite number of counterexamples can possibly occur. Otherwise, it is possible that counterexamples could continue to be reported forever and that the static analysis never reaches soundness. For example, suppose we try to construct an eventually sound interval analysis for a program $x\leftarrow m()$ with an integral variable $x$ by abstracting a set of counterexamples with the smallest interval that contains all the counterexamples. Such an analysis is {\it not} eventually sound. On the other hand, an analysis that abstracts counterexamples with $(-\infty,\infty)$ is sound and therefore
(vacuously) eventually sound. Finally, the former analysis is eventually sound (but not precise)
if after $N$ counterexamples the analysis outputs $(-\infty,\infty)$.

Also, it is permissible for a counterexample to simply never occur in any execution, e.g., in the above code, if the call to \texttts{randInt} in \texttts{foo} always returns \texttts{1}, then the counterexample $\texttts{y}\hookrightarrow o$ never occurs during any execution. However, eventual soundness is still satisfied since soundness is defined relative to the sequence of observed executions: if a counterexample exists but is never observed, then the analysis is still sound for all executions that are observed.


\subsection{Na\"{i}ve Algorithm}

We first describe a na\"{i}ve eventually sound points-to analysis. Recall that we cannot compute points-to edges for variables in missing code---our analysis only computes edges for visible variables in the program.

\paragraph{Optimistic analysis.}

We use the static analysis in Figure~\ref{fig:pointsto} to compute static points-to edges $\Pi$, assuming that calls to library functions are no-ops---i.e., the set of counterexamples is initially empty, i.e., $\Pi_{\text{miss}}\gets\emptyset$.

\paragraph{Runtime checks.}

A \emph{monitor} is instrumentation added to a statement $x\gets*$ (where $*$ stands for any valid subexpression). After executing this statement, the monitor issues a \emph{report} $(x\gets*,\bar{o})$, i.e., it records the value of the concrete object $\bar{o}$ pointed to by $x$ after executing $x\gets*$. A \emph{monitoring scheme} $M$ is a set of statements in the program to be monitored. Our goal is to design monitoring schemes that satisfy the following property:
\begin{definition}
\rm
We say a monitoring scheme $M$ is \emph{sound} if (i) for any execution where some counterexamples occur, $M$ reports one of them, and (ii) $M$ only reports counterexamples, i.e., it does not report false positives.
\end{definition}

Na\"{i}vely, it is sound to monitor every variable $x\in\mathcal{V}_P$. Then, we can map each concrete object $\bar{o}$ to its allocation:
\begin{definition}
\rm
An \emph{abstract object mapping} for an execution is a mapping $\bar{o}\leadsto o$, where $\bar{o}$ is a concrete object allocated at abstract object $o$.
\end{definition}
For every report $(x\gets X(),\bar{o})$, we add $\bar{o}\leadsto o$ to our abstract object mapping, where $o=(x\gets X())$. Then, for every report $(x\gets*,\bar{o})$ and $\bar{o}\leadsto o$, we conclude that $x\hookrightarrow o$ occurred dynamically, and if missing, report it as a counterexample. In our example, we monitor \texttts{libStr}, detect that $\bar{o}_{\text{str}}\leadsto o_{\text{str}}$, and report counterexamples $\texttts{data}\hookrightarrow\bar{o}_{\text{str}}$ and (if \texttts{randBool} returns true) $\texttts{dataCopy}\hookrightarrow\bar{o}_{\text{str}}$.

\paragraph{Updating the static analysis.}

We add every reported counterexample to $\Pi_{\text{miss}}$. Our static analysis in Figure~\ref{fig:pointsto} adds $\Pi_{\text{miss}}$ to $\Pi$ and computes the consequences of these added edges. Continuing our example, assuming \texttts{randBool} returns false, and only $\texttts{data}\hookrightarrow o_{\text{str}}$ is reported. Our static analysis adds $\texttts{data}\hookrightarrow o_{\text{str}}$ to $\Pi$ (rule 4) as well as its consequence $\texttts{dataCopy}\hookrightarrow o_{\text{str}}$ (rule 2).

\paragraph{Guarantees.}
Let $\Pi^*$ be the points-to edges computed using $\Pi_{\text{miss}}=\Pi_{\text{miss}}^*$, where $\Pi_{\text{miss}}^*$ is the set of all missing points-to edges. Then, our analysis is:
\begin{itemize}
\item {\bf Eventually sound:} Since $\Pi_{\text{miss}}^*$ is finite, only finitely many counterexamples are ever reported. Therefore, $\Pi$ is sound for all executions following the last reported counterexample.
\item {\bf Precise:} Any sound set of points-to edges $\Pi'$ must contain the missing points-to edges $\Pi_{\text{miss}}^*$. Therefore, $\Pi_{\text{miss}}\subseteq\Pi_{\text{miss}}^*\subseteq\Pi'$, so $\Pi\subseteq\Pi^*\subseteq\Pi'$.
\item {\bf Automatic:} Our static analysis requires no human input.
\end{itemize}
In our example, the computed static points-to set $\Pi$ is sound once the counterexample $\texttts{data}\hookrightarrow o_{\text{str}}$ is reported, since the static analysis then computes the remaining missing points-to edge $\texttts{dataCopy}\hookrightarrow o_{\text{str}}$.

\begin{figure*}[t]
\centering
\scriptsize
\begin{tabular}{cc}
\hline \\
{\bf Data Structure} & {\bf Rules for Construction} \\ \\
\hline \\
{\bf monitors} &
(allocation)
$M_{\text{alloc}}=\mathcal{O}_P$
\hspace{0.25in}
(function call)
$\dfrac{x\gets m(y)}{(x\gets m(y))\in M_{\text{call}}}$ \\
\multirow{2}{*}{{\Large $\downarrow$}} & \\
\hline \\
{\bf reports} &
(allocation)
$\dfrac{x\gets X(),\sss x\hookrightarrow\bar{o}}{(x\gets X(),\bar{o})\in R_{\text{alloc}}}$
\hspace{0.25in}
(function call)
$\dfrac{x\gets m(y),\sss x\hookrightarrow\bar{o}}{(x\gets m(y),\bar{o})\in R_{\text{call}}}$ \\
\multirow{2}{*}{{\Large $\downarrow$}} & \\
\hline \\
\multirow{4}{*}{$\begin{array}{c}\textbf{abstract object}\\\textbf{mapping}\end{array}$} &
(program abstract objects) $\dfrac{(x\gets X(),\bar{o})\in R_{\text{alloc}}}{\bar{o}\leadsto o=(x\gets X())}$ \\ \\
& (proxy objects) $\dfrac{(*,\bar{o})\not\in R_{\text{alloc}},\sss(x_1\gets m_1(y_1),\bar{o})\in R_{\text{call}},\sss...}{\bar{o}\leadsto p=\{m_1,...\}}$ \\
\multirow{2}{*}{{\Large $\downarrow$}} & \\
\hline \\
$\begin{array}{c}\textbf{missing}\\\textbf{points-to edges}\end{array}$ &
$\dfrac{(x\gets m(y),\bar{o})\in R_{\text{call}},\sss\bar{o}\leadsto o}{x\hookrightarrow o\in\Pi_{\text{miss}}}$ \\
\multirow{2}{*}{{\Large $\downarrow$}} & \\
\hline \\
$\begin{array}{c}\textbf{optimistic static}\\\textbf{points-to edges}\end{array}$ &
$\Pi=(\text{apply Figure~\ref{fig:pointsto} with the constructed }\Pi_{\text{miss}})$ \\ \\
\hline \\
\end{tabular}
\caption{Given a program $P$, \toolname adds monitors to $P$. It uses reports issued by these monitors during executions to compute the counterexamples $\Pi_{\text{miss}}$, which the static analysis in Figure~\ref{fig:pointsto} uses to compute optimistic points-to edges $\Pi\subseteq\mathcal{V}_P\times(\mathcal{O}_P\cup\mathcal{P})$.}
\label{fig:algo}
\end{figure*}

\subsection{Optimized Monitoring}
\label{sec:optimizedmonitoring}

We now describe how to reduce monitoring.

\paragraph{Restricting to function calls.}

Recall that monitoring \texttts{dataCopy} is unnecessary---the missing points-to edge $\texttts{dataCopy}\hookrightarrow o_{\text{str}}$ is computed by the static analysis once $\texttts{data}\hookrightarrow o_{\text{str}}$ is reported, so it suffices to monitor \texttts{data}. In general, it suffices to monitor function calls and allocations:
\begin{proposition}
\label{prop:sufficient}
\rm
The monitoring scheme $M_{\text{min}}=M_{\text{alloc}}\cup M_{\text{call}}$ is sound, where $M_{\text{alloc}}=\mathcal{O}$ and $M_{\text{call}}=\{x\gets m(y)\mid m\in\mathcal{M}\}$.
\end{proposition}
We give a proof in Appendix~\ref{sec:sufficientproof}. Figure~\ref{fig:algo} shows our algorithm using $M_{\text{min}}$.

\paragraph{Restricting to leaked abstract objects.}
We can further reduce the size of $M_{\text{alloc}}$. In particular, library functions can only access abstract objects reachable from the parameter $y$ of a call $x\gets m(y)$, which implies that the return value $r_m$ can only point to such an abstract object $o$. Thus, it suffices to restrict $M_{\text{alloc}}$ to include abstract objects that may leak into missing code.

In fact, it is even sound to use the monitoring scheme $\tilde{M}_{\text{alloc}}$, which monitors allocation statements $o$ such that $o$ may be explicitly passed to the library via a function call $x\gets m(y)$, where $y\hookrightarrow o$:
\begin{align*}
\tilde{M}_{\text{alloc}}=\{o\in\mathcal{O}\mid y\hookrightarrow o\in\Pi\text{ where }x\gets m(y)\}.
\end{align*}
This monitoring scheme is subtler than the schemes described previously, since the monitors $\tilde{M}_{\text{alloc}}$ depend on the current points-to edges $\Pi$. Therefore, the instrumentation may need to be updated when counterexamples are reported and $\Pi$ is updated. In particular, if $y\hookrightarrow o$ is newly added to $\Pi$, where $x\gets m(y)$, then $o$ is added to $\tilde{M}_{\text{alloc}}$ so the instrumentation must be updated.

We can soundly use $\tilde{M}_{\text{min}}=\tilde{M}_{\text{alloc}}\cup M_{\text{call}}$:
\begin{proposition}
\label{prop:betteralloc}
\rm
Let $\tilde{M}_{\text{min}}$ be constructed using the current points-to edges $\Pi$, and suppose the program is instrumented using $\tilde{M}_{\text{min}}$. If any counterexample occurs during execution, then some counterexample is reported.
\end{proposition}
We give a proof in Appendix~\ref{sec:betterallocproof}. Since the number of possible counterexamples is still finite, at some point no further counterexamples are reported. By Proposition~\ref{prop:betteralloc}, no counterexamples occur in any subsequent executions, i.e., $\Pi$ is sound for all subsequent executions.

\paragraph{Minimality}
Our monitoring scheme is minimal in the following sense:
\begin{proposition}
\rm
\label{prop:minimal}
Assume that the rules used to compute $\tilde{M}_{\text{alloc}}$ do not generate any false positives, i.e., for every allocation $o\in\tilde{M}_{\text{alloc}}$, there exists an execution during which a concrete object allocated at $o$ is passed as an argument to a library function. Then, for any strict subset $M\subsetneq\tilde{M}_{\text{min}}$, there exist implementations of the library and program executions such that $M$ fails to report a counterexample, i.e., using $M$ is not eventually sound.
\end{proposition}
In other words, our monitoring scheme is minimal except for potential imprecision when computing $\tilde{M}_{\text{alloc}}$. We give a proof in Appendix~\ref{sec:minimalproof}.

\section{Abstract Objects in the Library}
\label{sec:proxy}

We now remove the assumption that no allocations occur inside missing code.

\subsection{Proxy Objects}

Suppose that allocations can occur inside library code. Let $\mathcal{O}=\mathcal{O}_P\cup\mathcal{O}_L$, where abstract objects in $\mathcal{O}_P$ are in available code and abstract objects in $\mathcal{O}_L$ are in missing code. Then, our analysis cannot compute points-to edges $x\hookrightarrow o$, where $o\in\mathcal{O}_L$. As described previously, we assume that the static analysis only needs to compute relations involving program values. However, points-to edges $x\hookrightarrow o\in\mathcal{V}_P\times\mathcal{O}_L$ (i.e., $x$ is in the program but $o$ is not) are often needed to compute relations between program variables, e.g., aliasing and concrete types.

For example, in Figure~\ref{fig:main}, if \texttts{mkStr} is missing, then $o_{\text{str}}$ is missing, so our static analysis cannot compute $\texttts{str}\hookrightarrow o_{\text{str}}$ (among others). Furthermore, we do not assume the ability to instrument missing code, so we cannot dynamically detect these points-to edges. However, this points-to edge is needed to determine that $\texttts{str}$ may have type \texttts{String}, and that \texttts{str} and \texttts{data} may be aliased.

We handle allocations in library code by constructing the following:
\begin{definition}
\rm
A \emph{proxy object mapping} $\phi$ maps $\bar{o}\leadsto p$, where $\bar{o}$ is a concrete object allocated in missing code, and $p=\phi(\bar{o})\in\mathcal{P}$ is a fresh abstract object called a \emph{proxy object}; here, $\mathcal{P}$ is the set of all proxy objects.
\end{definition}
In other words, $\phi$ is the abstract object mapping for concrete objects allocated in missing code. We describe how to construct $\phi$ and $\mathcal{P}$ below.

Given $\phi$, our analysis proceeds as before. It makes optimistic assumptions, initializes $\Pi_{\text{miss}}\gets\emptyset$, and instruments the program using the monitoring scheme $\tilde{M}_{\text{min}}$ defined in Proposition~\ref{prop:sufficient}. For any report $(x\gets*,\bar{o})$, if $\bar{o}$ is not allocated at a visible allocation, our analysis concludes that $\bar{o}$ must have been allocated in missing code, so it adds $\bar{o}\leadsto p=\phi(\bar{o})$ to the abstract object mapping. Now, if a detected dynamic points-to edge $x\hookrightarrow p$ is missing, it is reported as a counterexample and added to $\Pi_{\text{miss}}\subseteq\mathcal{V}_P\times(\mathcal{O}_P\cup\mathcal{P})$, and $\Pi$ is recomputed using the static analysis in Figure~\ref{fig:pointsto}. If $\mathcal{P}$ is finite, then this approach is eventually sound, since there can only be a finite number of counterexamples $x\hookrightarrow p$.

In our example, \texttts{str} is monitored since \texttts{mkStr} is missing. Upon execution, our instrumentation detects $\texttts{str}\hookrightarrow\bar{o}_{\text{str}}$, and determines that $\bar{o}_{\text{str}}$ (allocated at $o_{\text{str}}$) is allocated in missing code. Supposing that $p_{\text{str}}=\phi(\bar{o}_{\text{str}})\in\mathcal{P}$, our analysis adds $\bar{o}_{\text{str}}\leadsto p_{\text{str}}$ to the abstract object mapping. Thus, if \texttts{randBool} returns false, our instrumentation reports counterexamples $\texttts{str}\hookrightarrow p_{\text{str}}$ and $\texttts{data}\hookrightarrow p_{\text{str}}$, which are added to $\Pi_{\text{miss}}$, from which our static analysis computes $\texttts{dataCopy}\hookrightarrow p_{\text{str}}\in\Pi$.

We now discuss how to construct $\phi$ and $\mathcal{P}$. The relevant information characterizing a concrete object is the following:
\begin{definition}
\rm
The \emph{dynamic footprint} of a concrete object $\bar{o}$ is the set of all visible variables that ever point to $\bar{o}$ during an execution.
\end{definition}
The concrete type of $\bar{o}$ may also be available to the static analysis, which we discuss in Section~\ref{sec:concretetypes}. Aside from concrete types, the dynamic footprint contains all information about $\bar{o}$ available to the static analysis, namely, the visible variables that point to $\bar{o}$. 

Then, the proxy object mapping $\phi$ should map each concrete object $\bar{o}$ to a proxy object $p$ so that the corresponding \emph{static footprint} $\{x\in\mathcal{V}_P\mid x\hookrightarrow p\in\Pi^*\}$ soundly overapproximates the dynamic footprint of $\bar{o}$ as precisely as possible. This way, clients of the points-to analysis can be eventually soundly and precisely computed (as long as they only depend on available information), e.g., it ensures that aliasing for program variables is eventually soundly and precisely computed (concrete types are eventually soundly and precisely computed using a simple extension; see Section~\ref{sec:concretetypes}).

On the other hand, $\phi$ should also avoid introducing unnecessary proxy objects, or else more executions may be required for the analysis to become sound. Two extremes highlight these opposing desirable properties:
\begin{itemize}
\item {\bf Unbounded $\mathcal{P}$:} Map each concrete object to a fresh proxy object $\phi(\bar{o})=p_{\bar{o}}$.
\item {\bf Singleton $\mathcal{P}$:} Map each concrete object to a single proxy object $\phi(\bar{o})=p$.
\end{itemize}
On the one hand, if we use a fresh proxy object for every concrete object, then there would be an unbounded number of proxy objects, which would mean our algorithm is no longer eventually sound (since there may be an unbounded number of missing points-to edges). Alternatively, using a single proxy object can be very imprecise; for example, for \emph{any} pair of calls $x\gets m(y)$ and $x'\gets m'(y')$, our analysis concludes that $x$ and $x'$ may be aliased.

We first describe an \emph{ideal proxy object mapping}, which constructs $\mathcal{P}$ as the set of possible dynamic footprints, and constructs $\phi$ to map $\bar{o}$ to its dynamic footprint. Points-to sets computed using \emph{any} static analysis together with the ideal proxy object mapping satisfy the above property, i.e., that the static footprints soundly overapproximate the dynamic footprints as precisely as possible.

Because the static analysis is flow-insensitive, the ideal proxy mapping is actually more precise than necessary. Therefore, our analysis uses a coarser proxy object mapping computed by our analysis, which essentially restricts the dynamic footprint to function return values. Finally, we show that this coarser proxy object mapping is as precise as the ideal proxy object mapping for our points-to analysis described in Figure~\ref{fig:pointsto}.

\subsection{Ideal Proxy Object Mapping}
\label{sec:idealproxyobject}

Our ``ideal'' construction of proxy objects exactly captures dynamic footprints:
\begin{definition}
\rm
An \emph{ideal proxy object} $\tilde{p}\in\tilde{\mathcal{P}}=2^{\mathcal{V}_P}$ is a set of visible variables. The \emph{ideal proxy object mapping} $\tilde{\phi}(\bar{o})\in\tilde{\mathcal{P}}$ is the dynamic footprint of $\bar{o}$.
\end{definition}
For a concrete object $\bar{o}$ allocated in missing code, we can compute $\tilde{\phi}(\bar{o})$ by monitoring all visible variables and identifying all visible variables that ever point to $\bar{o}$. In our example, suppose that the concrete object $\bar{o}_{\text{str}}$ is allocated at missing abstract object $o_{\text{str}}$ in an execution where \texttts{randBool} returns \texttts{false}. Then, $\tilde{\phi}$ maps $\bar{o}_{\text{str}}$ to ideal proxy object $\tilde{p}_{\text{str}}=\{\texttts{str},~\texttts{data}\}$. The reported counterexamples
\begin{align*}
\tilde{\Pi}_{\text{miss}}=\{\texttts{str}\hookrightarrow\tilde{p}_{\text{str}},~\texttts{data}\hookrightarrow\tilde{p}_{\text{str}}\}
\end{align*}
are added to our static analysis, which additionally computes $\texttts{dataCopy}\hookrightarrow\tilde{p}_{\text{str}}$.

Let $\tilde{\Pi}_{\text{miss}}^*\subseteq\mathcal{V}_P\times(\mathcal{O}_P\cup\tilde{\mathcal{P}})$ be the missing points to edges when using ideal proxy objects, and let $\tilde{\Pi}^*\subseteq\mathcal{V}_P\times(\mathcal{O}_P\cup\tilde{\mathcal{P}})$ be the points-to edges computed using $\Pi_{\text{miss}}=\tilde{\Pi}_{\text{miss}}^*$. Then:
\begin{proposition}
\label{prop:idealsoundness}
\rm
If $x\hookrightarrow\bar{o}$ occurs during execution and $\bar{o}$ is allocated at abstract object $o$, then $x\hookrightarrow o\in\tilde{\Pi}^*$ (if $o\in\mathcal{O}_P$) or $x\hookrightarrow\tilde{p}\in\tilde{\Pi}^*$ (where $\tilde{p}=\tilde{\phi}(\bar{o})$).
\end{proposition}
In other words, clients of the points-to analysis that only refer to program variables are eventually sound. For example, if two program variables $x$ and $y$ may be aliased, then there must be some execution in which they both point to a concrete object $\bar{o}$. Then, our analysis finds points-to edges $x\hookrightarrow\tilde{p}$ and $y\hookrightarrow\tilde{p}$, where $\tilde{p}=\tilde{\phi}(\bar{o})$, so the alias analysis determines that $x$ and $y$ may be aliased. Also:
\begin{proposition}
\label{prop:idealprecision}
\rm
Let $\Pi\subseteq\mathcal{V}_P\times(\mathcal{O}_P\cup\mathcal{O}_L)$ be the points-to set computed using the static analysis in Figure~\ref{fig:pointsto} with all code available (and $\Pi_{\text{miss}}=\emptyset$). For $o\in\mathcal{O}_P$, if $x\hookrightarrow o\in\tilde{\Pi}^*$, then $x\hookrightarrow o\in\Pi$. For $\tilde{p}=\tilde{\phi}(\bar{o})\in\tilde{\mathcal{P}}$, if $x\hookrightarrow\tilde{o}\in\Pi^*$, then $x\hookrightarrow o\in\Pi$, where $o$ is the statement where $\bar{o}$ was allocated.
\end{proposition}
In other words, $\tilde{\Pi}^*$ is at least as precise as the points-to edges $\Pi$ computed with all code available. We prove these two propositions in Appendix~\ref{sec:idealprecisionproof}. In our example, with all code available, we compute $\texttts{str}\hookrightarrow o_{\text{str}}$, $\texttts{data}\hookrightarrow o_{\text{str}}$, and $\texttts{dataCopy}\hookrightarrow o_{\text{str}}$, which is equivalent to $\tilde{\Pi}^*$ (replacing $\tilde{p}_{\text{str}}$ with $o_{\text{str}}$).

\subsection{Proxy Object Mapping}

The ideal proxy object mapping is more precise than necessary. Continuing our example, consider a second execution where \texttts{randBool} returns true. Then, the concrete object $\bar{o}_{\text{str}}'$ allocated at missing abstract object $o_{\text{str}}$ is mapped to the ideal proxy object $\tilde{p}_{\text{str}}'=~\{\texttts{str},~\texttts{data},~\texttts{dataCopy}\}$.

However, the static footprint of $\tilde{p}'$ equals that of $\tilde{p}$ (from the first execution, where \texttts{randBool} returns false), even though $\tilde{p}\not=\tilde{p}'$---i.e., $\bar{o}_{\text{str}}$ and $\bar{o}_{\text{str}}'$ map to different ideal proxy objects, but their relevant points-to behaviors appear identical to the (flow-insensitive) static analysis. In fact, all information about a concrete object available to the static analysis can be summarized by the following:
\begin{definition}
\rm
The \emph{dynamic function footprint} of a concrete object $\bar{o}$ is the set of library functions $m\in\mathcal{M}$ such that $r_m\hookrightarrow\bar{o}$ during execution.
\end{definition}
Now, we use the following proxy object mapping:
\begin{definition}
\rm
A \emph{proxy object} $p\in\mathcal{P}=2^{\mathcal{M}}$ is a set of library functions. The \emph{proxy object mapping} $\phi(\bar{o})\in\mathcal{P}$ is the dynamic function footprint of $\bar{o}$.
\end{definition}
To compute $\phi$, it suffices to monitor calls $x\gets m(y)$ to missing functions. Continuing our example, $\phi$ maps the concrete object $\bar{o}_{\text{str}}$ allocated at missing abstract object $o_{\text{str}}$ to $p_{\text{str}}=\{\texttts{mkStr},\texttts{get}\}$ regardless of the return value of \texttts{randBool}. If \texttts{randBool} returns true, then
\begin{align*}
\Pi_{\text{miss}}=\{\texttts{str}\hookrightarrow p_{\text{str}},~\texttts{data}\hookrightarrow p_{\text{str}},~\texttts{data}\hookrightarrow p_{\text{str}}\},
\end{align*}
in which case our static points-to analysis does not compute any additional edges. If \texttts{randBool} returns false, then
\begin{align*}
\Pi_{\text{miss}}=\{\texttts{str}\hookrightarrow p_{\text{str}},~\texttts{data}\hookrightarrow p_{\text{str}}\},
\end{align*}
from which our static analysis also computes $\texttts{dataCopy}\hookrightarrow p_{\text{str}}$. The static footprint of $p_{\text{str}}$ is the same either way, and also equals those of $\tilde{p}_{\text{str}}$ and $\tilde{p}_{\text{str}}'$.

Let $\Pi_{\text{miss}}^*\subseteq\mathcal{V}_P\times(\mathcal{O}_P\cup\mathcal{P})$ be the set of all missing points-to edges using proxy objects objects, and let $\Pi^*\subseteq\mathcal{V}_P\times(\mathcal{O}_P\cup\mathcal{P})$ be the points-to edges computed using $\Pi_{\text{miss}}=\Pi_{\text{miss}}^*$. Then:
\begin{proposition}
\label{prop:functionfootprint}
\rm
For any abstract object $o\in\mathcal{O}_P$, $x\hookrightarrow o\in\tilde{\Pi}^*\Leftrightarrow x\hookrightarrow o\in\Pi^*$. Furthermore, for any concrete object $\bar{o}$ allocated in missing code, letting $\tilde{p}=\tilde{\phi}(\bar{o})$ and $p=\phi(\bar{o})$, $x\hookrightarrow\tilde{p}\in\tilde{\Pi}^*\Leftrightarrow x\hookrightarrow p\in\Pi^*$.
\end{proposition}
In other words, the points-to edges computed using our proxy object mapping is as sound and precise as using the ideal proxy object mapping. Therefore, using proxy objects is also sound and precise in the sense of Propositions~\ref{prop:idealsoundness} and~\ref{prop:idealprecision}. We prove this proposition in Appendix~\ref{sec:functionfootprintproof}. 

Finally, the following result says that the monitoring scheme described in Section~\ref{sec:optimizedmonitoring} is still sound (it follows since we can compute $\phi$ using only $M_{\text{call}}$):
\begin{proposition}
\rm
The monitoring scheme $\tilde{M}_{\text{min}}$ is sound.
\end{proposition}

\section{Specification Inference}
\label{sec:infer}

Rather than simply adding reported missing points-to edges to $\Pi_{\text{miss}}$, we can use them to infer specifications summarizing missing code, which transfers information learned from the counterexample to other calls to the same library function.
We use the specification inference algorithm in~\cite{bastani2015specification}. Given a reported missing points-to edge $x\hookrightarrow o$, this algorithm infers specifications in two steps:
\begin{itemize}
\item {\bf Pessimistic assumptions:} Take $\hat{m}=m_{\text{pess}}$ for every missing function $m\in\mathcal{M}$, for some function $m_{\text{pess}}$ (see below), and run the static analysis using $\hat{m}$ in place of $m$.
\item {\bf Minimal statements:} Compute a minimal subset of \emph{pessimistic statements} (i.e., statements in the functions $m_{\text{pess}}$) that are needed to compute $x\hookrightarrow o$ statically; these statements are the inferred specifications.
\end{itemize}
The second step involves computing the static analysis using a shortest-path style algorithm. When computing the transitive closure according to the rules in Figure~\ref{fig:pointsto}, a priority queue is used in place of a worklist, where the priority of each points-to edge in the queue is the number of pessimistic statements needed to derive it. In particular, each time a rule is applied in conjunction with a pessimistic statement, the priority of the derived points-to edge is one more than the sum of the priorities of points-to edges in the premise.

\begin{figure}[t]
\begin{minipage}{0.45\textwidth}
\begin{scriptsize}
\begin{verbatim}
Object m_gen(Object ob) {
  while(true) {
    ob.f = ob; ob = ob.f; }
  return ob; }
\end{verbatim}
\end{scriptsize}
\end{minipage}
\begin{minipage}{0.45\textwidth}
\begin{scriptsize}
\begin{verbatim}
Object m_res(Object ob) {
  Object r;
  this.g = ob; r = ob; r = this; r = this.g;
  return r; }
\end{verbatim}
\end{scriptsize}
\end{minipage}
\caption{Pessimistic functions used for specification inference; $m_{\text{gen}}$ (left) is fully general (assuming functions do not access global state), whereas $m_{\text{res}}$ (right) is restricted to accessing only fields of the receiver. For simplicity, we omit the receiver in $m_{\text{gen}}$.}
\label{fig:pessimisticfunction}
\end{figure}

\paragraph{Pessimistic function.}
A key design choice is the pessimistic function $m_{\text{pess}}$ to use. The choice in~\cite{bastani2015specification}, which we term the \emph{general} function $m_{\text{gen}}$, is shown in Figure~\ref{fig:pessimisticfunction} (left). Using $m_{\text{gen}}$ is sound assuming library functions do not access global fields or allocate objects. However, it results in a huge search space of candidate specifications, so the inference algorithm produces many incorrect specifications. Instead, we use pessimistic assumptions that restrict the search space to only consider candidate specifications that are common in practice, in particular, that (i) do not accesses deep field paths, (ii) only access receiver fields, and (iii) assume the receiver has a single field $g$. These constraints lead to the \emph{restricted} function shown in Figure~\ref{fig:pessimisticfunction} (right).

\paragraph{Proxy object specifications.}
We separately infer \emph{proxy object specifications} of the form $(X,\{m\})$, where $X\in\mathcal{C}$ and $m\in\mathcal{M}$ is a library function. This specification says that a new object of type $X$ is allocated onto the return value of a function. We infer a proxy object specification for any proxy object $p\in\mathcal{P}$ we observe dynamically such that the function footprint of $p$ consists of a single function $m$.

\section{Extensions}
\label{sec:extension}

\subsection{Shared Fields}
\label{sec:sharedfields}

In Section~\ref{sec:assumptions}, we made the assumption that no shared fields $f\in\mathcal{F}_P\cap\mathcal{F}_L$ exist. Our analysis handles a shared field $f$ by converting stores $x.f\gets y$ and loads $x\gets y.f$ in the program into calls to setter and getter functions, respectively. To do so, we have to know which fields may be accessed by the library. We make the weaker assumption that the library does not access fields defined in the program---then, our analysis performs this conversion for every field $f$ defined in the library that is accessed by the program.

\subsection{Callbacks}
\label{sec:callbacks}

Android apps can register callbacks to be invoked by Android when certain events occur, e.g., the program can implement the callback \texttts{onLocationChanged}, which is invoked when the user location changes. If callbacks are not specified, then the static analysis may unsoundly mark them as unreachable. We use the approach in~\cite{bastani2015interactively} to eventually soundly compute reachable program functions. In particular, a \emph{potential callback}, is a program function that overrides a framework function. Intuitively, potential callbacks are the functions ``known'' to the framework. For each potential callback $m$ that is marked as unreachable by the static analysis, we instrument $m$ to record whether $m$ is ever reached. This algorithm is eventually sound since there are only finitely many potential callbacks. Also, the instrumentation eventually incurs no overhead---once no more counterexamples are reported, the instrumentation is never triggered.

In addition, some callbacks are passed parameters from the Android framework. For example, consider the code on the left:

\vspace{0.05in}

\begin{minipage}{0.45\textwidth}
\begin{scriptsize}
\begin{verbatim}
void onLocationChanged(Location loc) {
  Location copy = loc; }
\end{verbatim}
\end{scriptsize}
\end{minipage}
\begin{minipage}{0.45\textwidth}
\begin{scriptsize}
\begin{verbatim}
void onLocationChanged() {
  Location loc = Location.getLocation();
  Location copy = loc; }
\end{verbatim}
\end{scriptsize}
\end{minipage}

\vspace{0.05in}

Here, \texttts{loc} points to an abstract object $o_{\text{loc}}$. In this case, $o_{\text{loc}}$ is allocated in the framework, but it may also be allocated in program code. We must specify the abstract objects that \texttts{loc} may point to, or else our points-to analysis is unsound. The code on the right replaces the parameter with a call that retrieves \texttt{loc} from the framework, which is semantically equivalent to the code on the left. Thus, we can think of \texttts{loc} as a ``return value'' passed to \texttts{onLocationChanged}; by Proposition~\ref{prop:sufficient}, it suffices to monitor all callback parameters.

\subsection{Concrete Types}
\label{sec:concretetypes}

Some client analyses additionally need the concrete type $X\in\mathcal{C}$ of abstract objects $o=(x\gets X())$, for example, virtual call resolution. To compute concrete types for proxy objects, each monitor $x\gets*$ additionally records the concrete type of the concrete object pointed to by $x$ after executing the statement. Then, the proxy objects are extended to $\mathcal{P}=\mathcal{C}\times2^{\mathcal{M}}$, and the proxy object mapping $\phi$ maps $\bar{o}\leadsto p=(X,F)$, where $F\subseteq2^{\mathcal{M}}$ is the dynamic function footprint of $\bar{o}$, and $X\in\mathcal{C}$ is the recorded concrete type of $\bar{o}$.

\subsection{Context- and Object-Sensitivity}
\label{sec:contextsensitivity}

Our analysis extends to $k$-context-sensitive points-to analyses with two changes. First, the abstract objects considered are typically pairs $o=(c,h)$, where $c$ is a calling context and $h$ is an allocation statement, so monitors on allocation statements $x\gets X()$ also record the top $k$ elements of the current callstack. Second, the points-to edge typically keeps track of the calling context $d$ in which a variable $v$ may point to abstract object $o$. Therefore, monitors on calls to missing functions $x\gets m(y)$ also record the top $k$ elements of the current callstack.

In particular, our analysis may (i) detect that $(d,v)\hookrightarrow\bar{o}$ (i.e., $d$ is the callstack when $v$ pointed to $\bar{o}$), and (ii) $\bar{o}\leadsto(c,h)$ (i.e., $\bar{o}$ was allocated at statement $h$, and $c$ is the callstack when $\bar{o}$ was allocated). Then, our analysis reports missing points-to edge $(d,v)\hookrightarrow(c,h)$. We use a 1-CFA points-to analysis in our evaluation; in this case, the calling context is simply the function in which the allocation or call to a missing function occurs. Our approach can be extended to handle object-sensitive analyses, by including instrumentation that records the calling context (which now includes the value of the receiver).

Finally, we can also handle on-the-fly callgraph construction---if a missing points-to edge $x\hookrightarrow o$ is reported, and there is a virtual function call $x.m()$ in the program, then the possible targets of $x.m()$ are updated to take into account the concrete type of $o$. The instrumentation may need to be updated based on this new information. Assuming the number of possible call targets is finite, this approach is eventually sound.

\section{Implementation}
\label{sec:implementation}

We have implemented our eventually sound points-to analysis, including all extensions described in Section~\ref{sec:extension} (using a 1-CFA points-to analysis), for Android apps in a tool called \toolname. The missing code consists of Android framework methods, which we assume cannot be statically analyzed (since the Android framework heavily uses native code and reflection) or instrumented (which requires a custom Android installation). The static analysis framework we use predates \toolname, and uses hand-written specifications to model missing code. Specifications have only been written for methods deemed relevant to a static information flow client---of the more than 4,000 Android framework classes, only 175 classes have specifications. Framework methods without specifications appear as missing code to our static analysis.

\toolname instruments Android apps using our optimized monitoring scheme $\tilde{M}_{\text{min}}$. It computes eventually sound points-to sets and infers specifications based on reported missing points-to edges. We instrument apps using the Smali assembler and disassembler~\cite{smali2015}. To monitor a statement \texttts{x=...}, we record (i) the value \texttts{System.identityHashCode(x)}, which uniquely identifies the concrete object pointed to by \texttts{x}, (ii) the concrete type \texttts{x.getClass()} of \texttts{x}, and (iii) the method containing the statement and the offset of that statement in the method. This data is uploaded to a server in batches (by default, once every 500ms), which post-processes it to compute missing points-to edges and infer specifications. To obtain traces, we execute apps in the Android emulator and use Monkey~\cite{google2016monkey} to inject touch events. We measure overhead using the Android profiler.

We have implemented the points-to analysis, the monitoring optimization, and the specification inference algorithm in a version of the Chord program analysis framework \cite{naik2006effective} modified to use Soot as a front end \cite{vallee1999soot}. The specification inference algorithm is based on shortest-path context-free reachability, described in~\cite{bastani2015specification}. We use a 1-CFA points-to analysis. As we discuss Section~\ref{sec:contextsensitivity}, using our more precise points-to analysis is eventually sound. Finally, we compute an information flow analysis based on this points-to analysis. The information flow analysis is standard---we look for paths from annotated sources (e.g., location, contacts, etc.) to annotated sinks (e.g., SMS messages, Internet, etc.) in the Android framework~\cite{fuchs2009scandroid,arzt2014flowdroid,bastani2015specification}. All analyses are computed using BDDBDDB~\cite{whaley2004cloning}.

\section{Evaluation}
\label{sec:experiments}

We evaluate \toolname on a benchmark of 73 Android apps, including battery monitors, games, wallpaper apps, and contact managers. These apps were obtained from a variety of sources, including malware from a major security company (primarily apps that leaked sensitive information such as location, contacts, SMS messages, etc.) and benign apps from Google Play Store. We omit 11 apps that fail to run on the standard Android emulator, leaving 62 apps. First, we use \toolname to instrument each Android app and study the instrumentation overhead. Second, we show how \toolname computes points-to edges over time, and show that the number of computed edges does not explode. Third, we show how our analysis can be used to improve an information flow client.

\subsection{Instrumentation Overhead}
\label{sec:overhead}

We evaluate the runtime overhead of our monitoring scheme $\tilde{M}_{\text{min}}$ described in Section~\ref{sec:optimizedmonitoring}. Recall from Section~\ref{sec:optimizedmonitoring} that our optimized instrumentation scheme may add instrumentation over time. We consider two settings:
\begin{itemize}
\item {\bf Initial:} This configuration represents the instrumentation overhead for a new app using the current program analysis. In particular, we use the initial instrumentation scheme where $\tilde{M}_{\text{alloc}}$ is constructed with no known counterexamples (i.e., $\Pi_{\text{miss}}=\emptyset$). Also, we use all existing handwritten points-to specifications, representing the realistic scenario where some manually provided information is used in addition to automatic inference.
\item {\bf Worst:} This configuration represents the absolute upper bound on the overhead. In particular, we monitor apps using the worst-case instrumentation scheme where $\tilde{M}_{\text{alloc}}$ contains all abstract objects that may leak into missing code. Furthermore, we remove all handwritten points-to specifications.
\end{itemize}
We executed instrumented apps in a standard emulator using Monkey for one hour, and then used our algorithm to compute points-to sets.

\paragraph{Results.}

We show the highest runtime overheads in Figure~\ref{fig:overhead} (left), including the runtime overhead from recording data and the amount of data generated in an hour, for both the initial setting and the worst-case setting.\footnote{We ran a small subset of apps on a real device and consistently measured smaller overhead; the emulator gives a coarser measure of execution time that we round up.} Columns ``updated'' and ``\# specs'' are discussed below. We plot the runtime overhead of our recording instrumentation in Figure~\ref{fig:tracefraction} (a), where the apps along the $x$-axis are sorted according to the overhead in the worst-case setting.

\paragraph{Discussion.}

The overhead incurred by recording data is less than 5\% for more than half of the apps, showing that in most cases the automatically instrumented programs have acceptable performance. Even in the worst case, more than half the apps have less than 10\% overhead. Still, there are outliers, with 5 apps incurring more than 20\% overhead with initial instrumentation, and in the worst-case, 9 apps incurred more than 20\% overhead. Unsurprisingly, the high-overhead outliers have instrumentation in inner loops of the app; in such cases the overhead can be reduced (see below). Finally, the amount of data generated is very small. Even in the worst case, for all but one of the apps, less than 1.0 MB of (compressed) data was generated in one hour. The median amount of data generated is about 2.0 KB, which is negligible. Data can therefore be stored and transmitted when the app is idle, so the overhead due to uploading data does not affect the user experience.

\begin{figure}[t]
\scriptsize
\begin{center}
\begin{tabular}{c|r|r|r|r|r|r}
\hline
\multirow{2}{*}{{\bf Rank}} & \multicolumn{4}{c|}{{\bf Recording Overhead (\%)}} & \multicolumn{2}{c}{{\bf Data (MB/hr)}} \\
\cline{2-7}
& initial & \multicolumn{1}{c|}{updated} & \multicolumn{1}{c|}{\# specs} & worst & initial & worst \\
\hline
1 & 50.0 & 31.0 & 15 & 91.6 & 0.71 & 1.72 \\
2 & 46.8 & 17.6 & 10 & 78.2 & 0.46 & 0.62 \\
3 & 39.2 & 6.7  & 5 & 76.9 & 0.41 & 0.51 \\
4 & 30.6 & 6.9  & 1 & 75.1 & 0.40 & 0.46 \\
5 & 28.3 & 19.7 & 5 & 74.8 & 0.37 & 0.38 \\
6 & 19.9 & \multicolumn{1}{c|}{--} & \multicolumn{1}{c|}{--} & 51.9 & 0.34 & 0.26 \\
\hline
{\bf median} & 4.3 & \multicolumn{1}{c|}{--} & \multicolumn{1}{c|}{--} & 8.6 & 0.02 & 0.02 \\
\hline
\end{tabular}
\end{center}
\caption{The runtime overhead from recording data and the (compressed) size of the data generated in one hour. Each is divided into initial and worst-case. The ``updated'' overhead is obtained by adding specifications to the system to reduce monitoring, and ``\# specs'' is the number of specifications added to do so. For each column, the table shows the largest six values and the median value across our benchmark.}
\label{fig:overhead}
\end{figure}

\paragraph{Reducing runtime overhead.}
Any program where instrumentation is required in a tight inner loop is particularly challenging for dynamic analysis. Standard sampling techniques can be used to reduce overhead in these cases~\cite{liblit2003bug}. Additionally, both $\tilde{M}_{\text{alloc}}$ and $M_{\text{call}}$ decrease in size as specifications are added and reach zero when there are no missing specifications. For a given program, we can test the program to determine which monitors are frequently triggered, and compute which missing functions require specifications for these monitors to be removed. Providing or inferring specifications for these functions would allow us to remove the expensive monitors. We do so for the five apps with initial overhead greater than 20\%. In Figure~\ref{fig:overhead} (left), we show both the number of specifications we added for that app (``\# spec'') and the resulting overhead (``updated''). For all but the top app, we were able to reduce the overhead below 20\% by adding specifications for at most 10 Android framework methods; again, the overhead can be reduced to any desired level by adding more specifications.

\subsection{Reported Counterexamples}

Next, we evaluate how the computed points-to edges vary over time. In particular, we show that the number of reported counterexamples does not explode over time---otherwise, the number of counterexamples discovered in production may be unacceptably high. Furthermore, we show that a tail of reported counterexamples continues to occur for some apps, which shows that running instrumented apps in production is necessary. This experiment uses the worst-case setting where all handwritten specifications have been removed.

\paragraph{Counterexamples over time.}

\begin{figure*}[t]
\centering
\begin{tabular}{cc}
\includegraphics[width=0.5\textwidth]{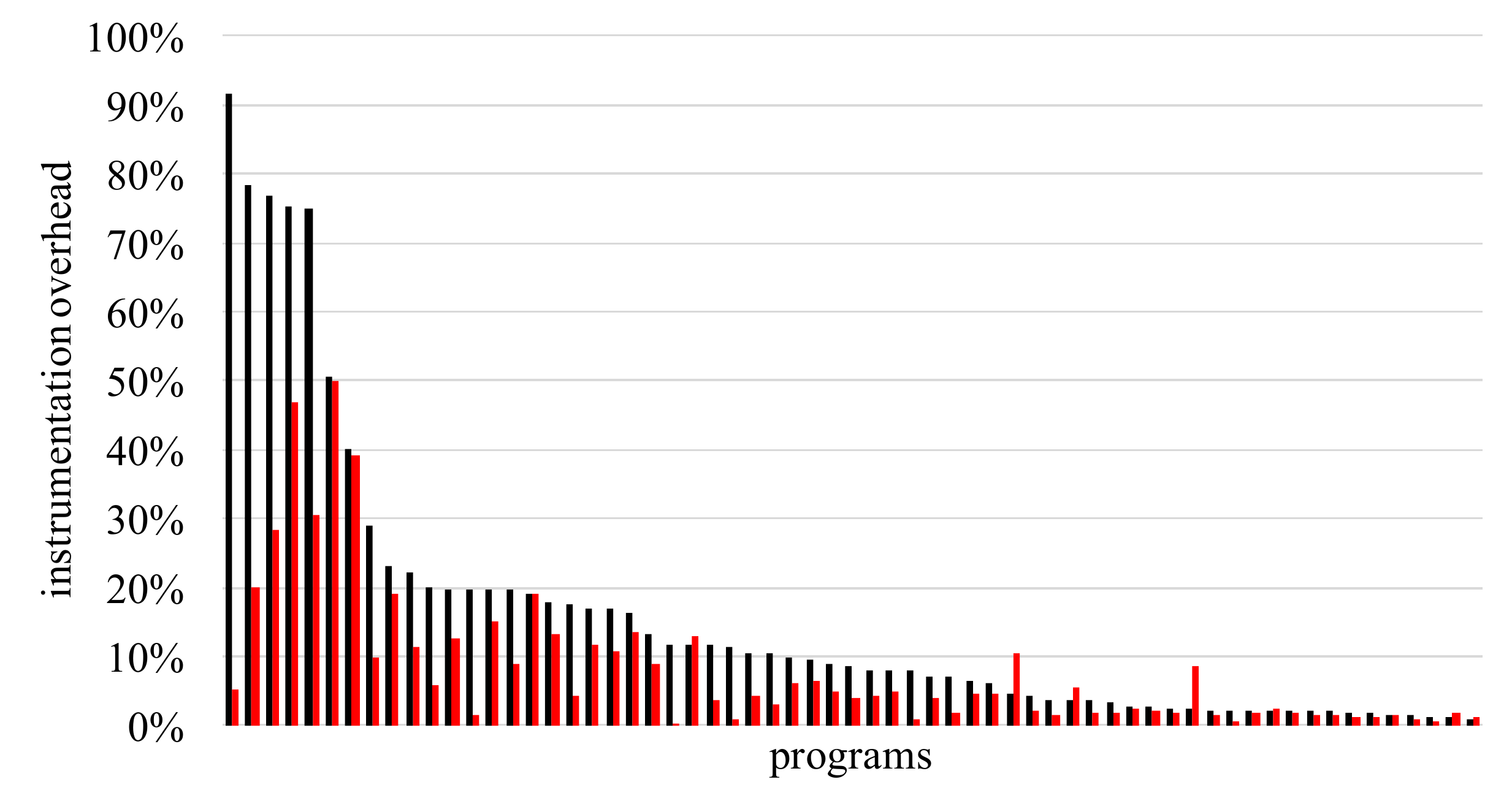}
& \includegraphics[width=0.5\textwidth]{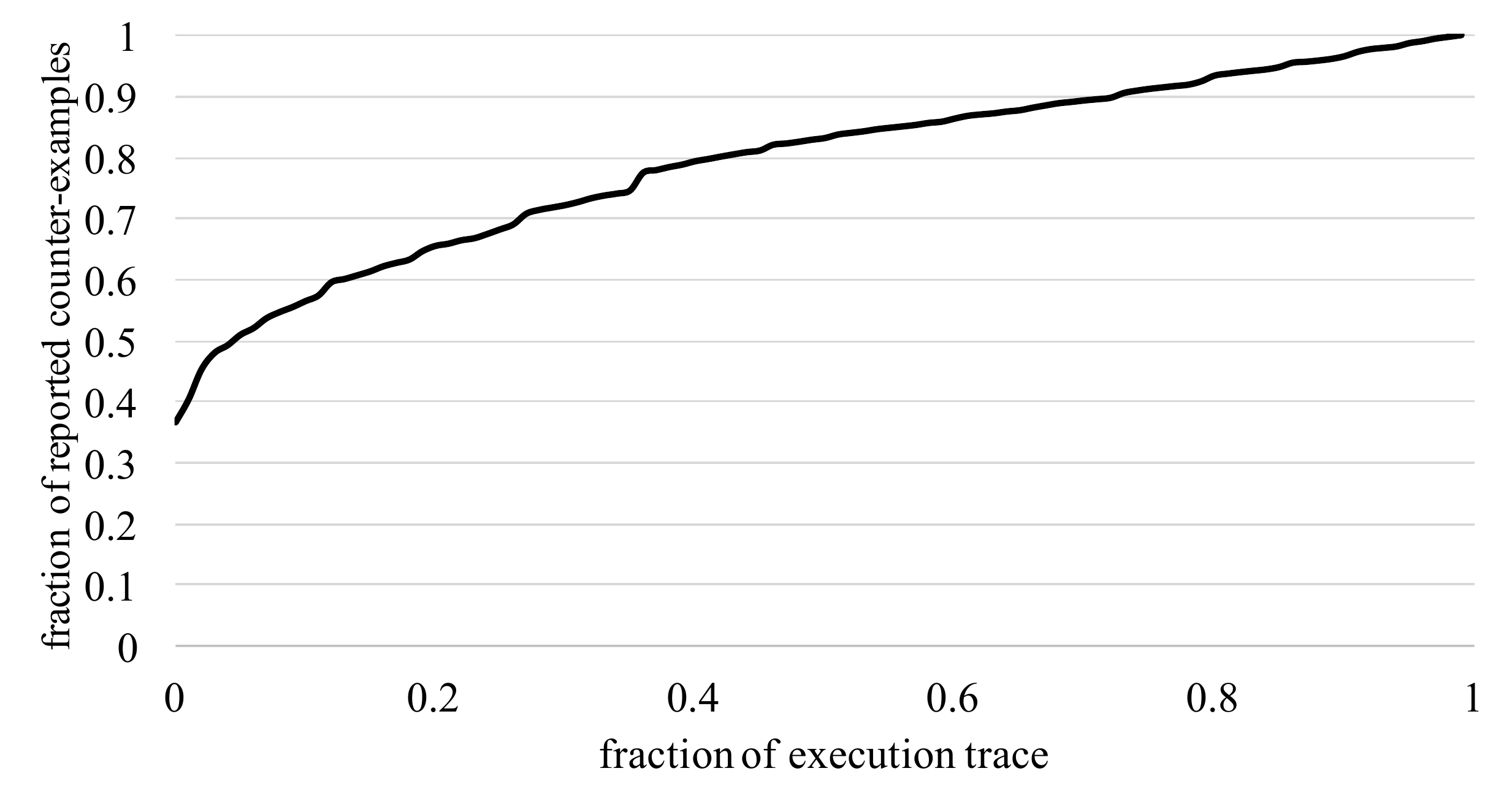} \\
(a) & (b) \\
\includegraphics[width=0.5\textwidth]{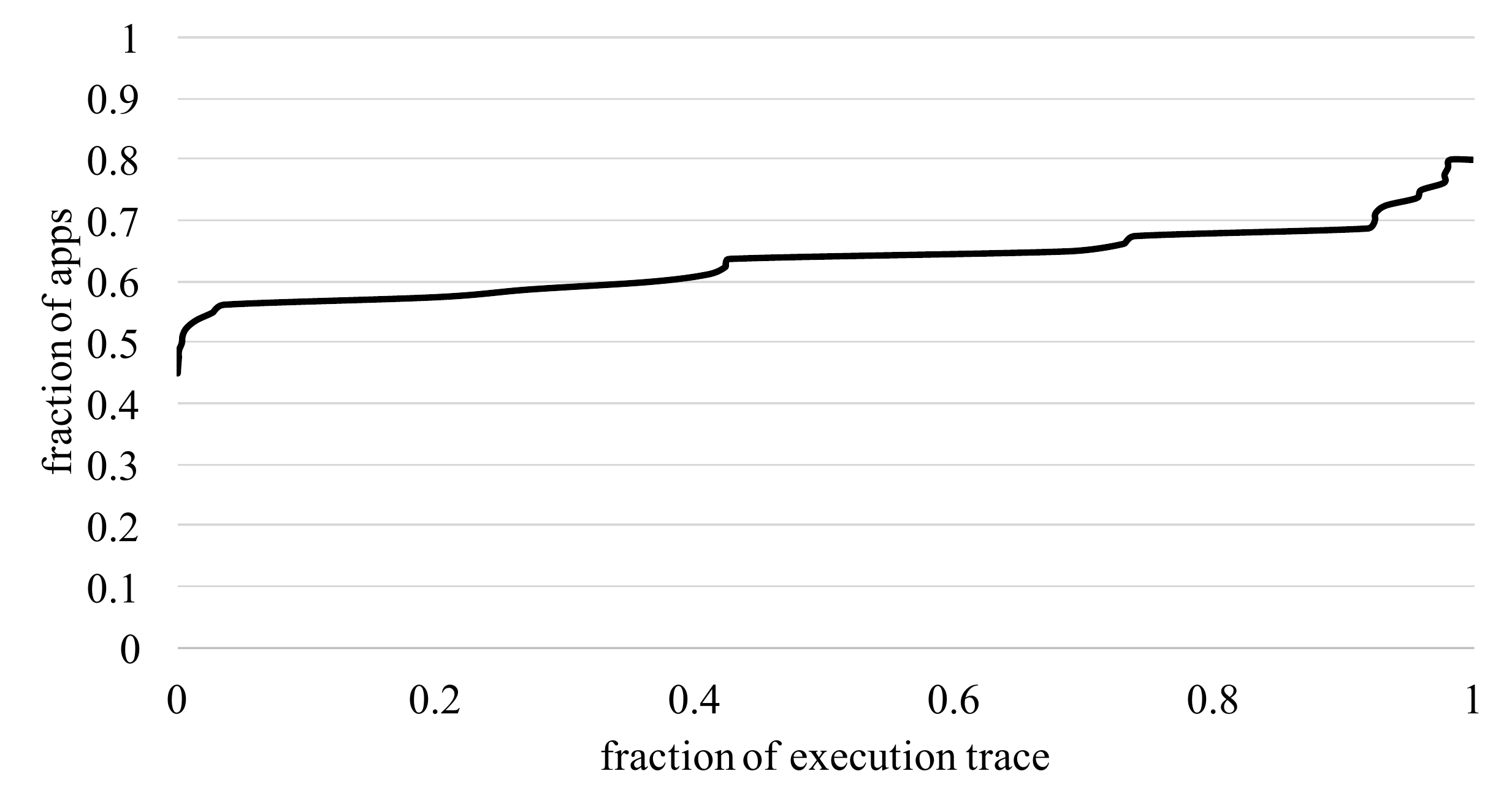}
& \includegraphics[width=0.5\textwidth]{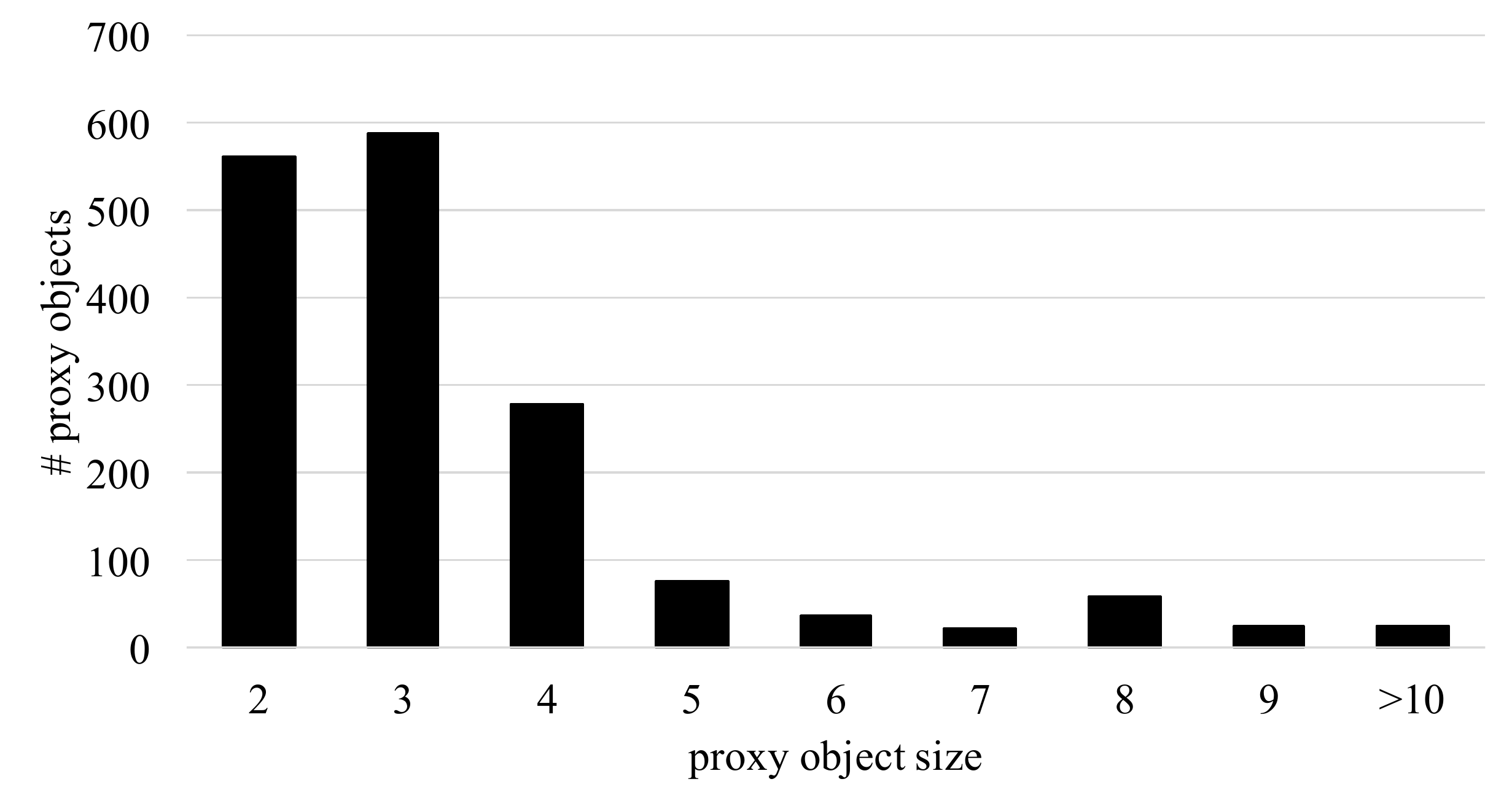} \\
(c) & (d)
\end{tabular}
\caption{(a) Runtime overhead of our recording instrumentation in the worst-case setting (black) and the initial setting (red). The overheads are sorted by the overhead for the worst-case setting. (b) The $x$-axis is a fraction of the execution trace, and the curve shows the fraction of discovered missing points-to edges that discovered up to that point in the execution trace (averaged over all apps). (c) The $x$-axis is again a fraction of the execution trace, and the curve shows the number of apps for which no further missing points-to edges are reported after that point in the execution trace. (c) The distribution of the sizes of the proxy objects (i.e., the size of its dynamic function footprint), omitting footprints of size one.}
\label{fig:tracefraction}
\end{figure*}

Figure~\ref{fig:tracefraction} (b) shows the cumulative number of reported missing points-to edges as execution progresses. More precisely, for each point in the execution trace ($x$-axis), it shows what fraction of reported missing points-to edges were discovered before that point. The values are averaged over all apps. By definition, at the end of the trace ($x=1.0$), the fraction of reported missing points-to edges also goes to $y=1.0$.

As can be seen, a large fraction of reports are made early on, with about 65\% of reports made within 20\% of the execution trace. We expect the number of reported counterexamples to continue to converge over time, and should not grow substantially larger. However, the curve is still increasing by the end of the execution trace, which indicates that more missing points-to edges are still being reported. Therefore, it is important to continue monitoring these apps in production to detect additional counterexamples.

\paragraph{Last discovered counterexample.}

Figure~\ref{fig:tracefraction} (c) shows the point in the execution during which the final reported missing points-to edge occurs. More precisely, for each point in the execution trace ($x$-axis), it shows the fraction of the apps for which the final reported missing points-to edge was reported before that point. This curve goes to $y=1.0$ at $x=1.0$, but we cut off apps that have reported missing points-to edges in the last 1\% of execution.

A large fraction of apps (about 45\%) report no counterexamples. About 10\% of apps report no further counterexamples after the first 5\% of the trace. At the opposite end of the trace, about 20\% of apps have the final reported counterexample in the last 1\% to 10\% of the trace, and 20\% have the final reported counterexample in the final 1\% of the trace, so more counterexamples likely remain. Again, this shows that we must continue to monitor apps in production.

\paragraph{Proxy object sizes.}

Since there are an exponential number of possible proxy objects (in the number of missing functions), we could hypothetically continue to discover many new proxy objects over time. In Figure~\ref{fig:tracefraction} (d), we show the sizes of the dynamic function footprints of the reported proxy objects. More precisely, we show the number of reported proxy objects ($y$-axis) for different dynamic function footprint sizes. As can be seen, the vast majority (85\%) of reported proxy objects have four or fewer functions in their dynamic function footprint. While there is a long tail of proxy objects with large function footprint sizes, there is no exponential blowup in the number of proxy objects discovered, ensuring that the analysis does not diverge due to proxy objects.

\subsection{Specification Inference and a Static Information Flow Client}

\begin{figure}[t]
\begin{minipage}{0.5\textwidth}
\scriptsize
\begin{center}
\begin{tabular}{c|r|r|r|r}
\hline
& {\bf Existing} & {\bf Inferred} & {\bf Correct} & {\bf Accuracy} \\
\hline
$m_{\text{res}}$ & 299 & 58 & 49 & 0.84 \\
$m_{\text{gen}}$ & 299 & 159 & 33 & 0.22 \\
proxy object & 330 & 422 & 383 & 0.91 \\
\hline
\end{tabular}
\end{center}
\end{minipage}
\hspace{0.2in}
\begin{minipage}{0.4\textwidth}
\scriptsize
\begin{center}
\begin{tabular}{c|r|r}
\hline
{\bf App} & {\bf Jimple LOC} & {\bf Time (min.)} \\
\hline
0C2B78 & 322K & 3.38 \\
b9ac05 & 268K & 1.06 \\
highrail & 247K & 1.49 \\
game & 174K & 0.08 \\
androng & 170K & 0.48 \\
\hline
median & 19K & 0.08 \\
\hline
\end{tabular}
\end{center}
\end{minipage}
\caption{Left: The number of specifications inferred using each $m_{\text{res}}$ and $m_{\text{gen}}$, and the number of proxy object specifications inferred. Right: Statistics for five largest apps used in our evaluation, including the number of Jimple lines of code (i.e., the intermediate representation used by Soot), and the running time of specification inference.}
\label{fig:specificationinference}
\end{figure}

Finally, we evaluate whether \toolname benefits an information flow client. We first infer specifications using the the algorithm in Section~\ref{sec:infer}, and then run our information flow client on various sets of specifications. The information flow analysis is standard---we look for paths from a set of annotated sources (e.g., location) to a set of annotated sinks (e.g., Internet) in the Android framework~\cite{fuchs2009scandroid,arzt2014flowdroid,feng2014apposcopy,bastani2015specification}. We demonstrate that the inferred specifications enables client to discover more information flows. However, many of the information flows remain undiscovered because the dynamic analysis is an underapproximation, which again motivates the need to run instrumented apps in production.

\paragraph{Specification inference.}

We remove all points-to specifications from \toolname, and then infer specifications from reported counterexamples. Figure~\ref{fig:specificationinference} (left) summarizes the inferred specifications---a specification is correct if it exactly equals the existing specification (or the one we would have written). Using $m_{\text{res}}$ is substantially more accurate than using $m_{\text{gen}}$, which does not infer a single additional specification. Compared to existing specifications, we inferred 174 new points-to specifications, of which 160 were proxy object specifications.

Furthermore, in Figure~\ref{fig:specificationinference} (right), we show statistics for the five largest apps in our benchmark along with the running time of the inference algorithm (the information flow analysis runs much faster than the inference algorithm). As can be seen, inference scales even to very large apps.

\paragraph{Static information flow client.}
In Figure~\ref{fig:malware}, we report the number of information flows and the number of malicious apps detected with varying sets of specifications (one malicious app can exhibit multiple flows). If we assume that all points-to specifications are missing (``Empty''), then the information flow client does not identify any information flows, whereas using inferred specifications (``Inf.'') computes a small number of flows.

A more representative use case is where the analysis has an incomplete baseline consisting of the most commonly used specifications (``Base''). Our baseline contains specifications for the essential Android framework classes \texttts{Bundle} and \texttts{Intent}, for the commonly used data serialization classes \texttts{JSONArray}, \texttts{JSONObject}, and \texttts{BasicNameValuePair}, and for a few methods in \texttts{java.util}. As can be seen from Figure~\ref{fig:malware}, when using the baseline in conjunction with the inferred specifications, the analysis computes a considerable number of additional flows compared to using the baseline alone (39 vs. 3). The reason the inferred specifications are more beneficial in this setting is that an information flow usually depends on multiple specifications---if a single one of these specifications is missing, then the flow is missing.

Compared to the existing, handwritten specifications (``Ex.''), using inferred specifications (together with the baseline specifications) identifies almost a third of the information flows. However, random testing cannot reveal all malicious behavior, since malware developers often try to hide malicious behaviors by triggering them only in response to very specific events, for example, at a certain time~\cite{mishra2007fixed}. Therefore, our instrumentation is necessary to ensure that we identify additional malicious behaviors as soon as or before they occur, thereby limiting potential damage. Note that we do not recover any new flows when combining inferred specifications with existing specifications---prior to our evaluation, we have already identified all specifications needed to recover flows in these apps.

Finally, an alternative way to evaluate the value of the inferred specifications, we consider omitting the inferred specifications from the set of inferred specifications. Doing so limits the information flow client to identify only 34 flows, which demonstrates that the inferred specifications are crucial for finding many of the information flows present in these apps.

\begin{figure}[t]
\begin{center}
\scriptsize
\begin{tabular}{c|r|r|r|r|r|r|r}
\hline
 & {\bf Empty} & {\bf Inf.} & {\bf Base} & {\bf Base $\cup$ Inf.} & {\bf Ex. $\setminus$ Inf.} & {\bf Ex.} & {\bf Ex. $\cup$ Inf.} \\
\hline
specs. & 0 & 432 & 189 & 621 & 371 & 629 & 803 \\
flows & 0 & 4 & 3 & 39 & 34 & 125 & 125 \\
malware & 0 & 2 & 3 & 22 & 12 & 49 & 49 \\
\hline
\end{tabular}
\end{center}
\caption{Comparison of different sets of specifications: ``Base'' includes the most frequently used specifications, ``Inf.'' includes the inferred specifications, and ``Ex.'' includes all handwritten specifications. For each set of specifications, we show the number of specifications in that set (``specs.''), the number of information flows computed using those specifications (``flows''), and the number of malicious apps identified, i.e., some malicious information flow was discovered (``malware'').}
\label{fig:malware}
\end{figure}

\section{Discussion}
\label{sec:discussion}

\paragraph{Dynamically loaded code.}

Our approach can be used for dynamically loaded code---the dynamically loaded code is taken to be the missing code, and the code that loads the dynamically loaded code is the available code. We guarantee eventual soundness for points-to edges in the available code. If points-to edges for dynamically loaded code must be computed, then the loaded code can be reported to the static analysis, but the analysis is no longer eventually sound---infinitely many reports may be issued since infinitely many different code fragments may be loaded.

\paragraph{Eventual soundness for clients.}

Our approach is automatically eventually sound for client analyses that depend only on aliasing information and concrete types for visible program variables (e.g., callgraph resolution). In general, missing code can introduce unsoundness into the static information flow analysis beyond missing points-to edges. For example, consider the code


\begin{center}
\begin{minipage}{0.45\textwidth}
\begin{scriptsize}
\begin{verbatim}
void main() { // program
  int val = source();
  int valDup = add(val, 1);
  sink(valDup); }
\end{verbatim}
\end{scriptsize}
\end{minipage}
\begin{minipage}{0.45\textwidth}
\begin{scriptsize}
\begin{verbatim}
int add(int x, int y) {  // library
  return x+y; }
\end{verbatim}
\end{scriptsize}
\end{minipage}
\end{center}


\noindent which calls the missing function \texttts{add}. Even with a sound points-to analysis, the static analysis would not recover the taint flow from \texttts{source} to \texttts{sink}. Sources and sinks in missing code must be specified, since there is no way to detect whether calling missing code leaks information out of the system or introduces sensitive information into the system.

In general, we can perform eventually sound analysis for clients that are abstract interpretations with finite abstract domain (at least, satisfying the ascending chain condition)~\cite{cousot1977abstract}, if the abstraction function $\alpha$ can be computed for values in the available code based on observations in the available code alone. In particular, for a call $y\gets m(x)$, the concrete values of $x$ and $y$ are recorded. Then, we can construct a transfer function $f_m$ to be analyzed in place of $m$. Initially, $\bot=f_m(\alpha(x))$ for all $x$; whenever a previously unobserved relation $\alpha(y)=f_m(\alpha(x))$ is detected during execution, a report is issued and $f_m$ updated. Since the abstract domain is finite, only finitely many reports can be issued, so the analysis is eventually sound. Finally, we use our points-to analysis to handle aliasing.

The challenge with information flow is that the abstraction function cannot be computed from observations in the available code alone, since information flow is a property of the computation, not just the input-output values. It may be possible to use techniques such as multi-execution~\cite{devriese2010noninterference}, which keep pair of values $\langle x_{\text{private}},x_{\text{public}}\rangle$ for each (visible) program variable $x$, where $x_{\text{private}}$ may depend on sensitive data whereas $x_{\text{public}}$ does not. For example, the value for program variable \texttts{val} may be $\langle14,0\rangle$, where $14$ is a sensitive value and $0$ is a public value. Then, we can execute \texttts{add} using both $\texttts{x}=14$ and $\texttts{x}=0$, and obtain return value $r_{\texttt{add}}=\langle15,1\rangle$. Since these two values differ, we conclude that $r_{\texttt{add}}$ depends on the sensitive input $14$, and report that \texttts{add} transfers information from its argument \texttts{x} to its return value $r_{\texttt{add}}$. Essentially, this approach transforms the program so the abstraction function becomes computable. Alternatively, existing techniques for specification inference such as~\cite{bastani2015specification} may be used to infer specifications describing how information flows through missing code.

\section{Related Work}
\label{sec:related}

\paragraph{Program monitoring.}

There has been work using runtime checks to complement static analysis. For instance, ~\cite{bodden2011taming} proposes to use dynamic information to resolve reflective call targets, and then instruments the program to report additional counterexamples. Similarly,~\cite{bastani2015interactively} proposes to compute reachable code by inserting runtime checks to report counterexamples to optimistic assumptions, and~\cite{flanagan2006hybrid} uses a combination of static type checking and runtime checks to enforce type safety. Points-to analysis with missing code is far more challenging, because dynamic points-to analysis incurs unreasonable overhead~\cite{mock2001dynamic,clapp2015modelgen}, and also requires instrumenting missing code.

Additionally,~\cite{hirzel2007fast} uses dynamic information to complement static points-to analysis. However, their analysis is unreasonably imprecise for programs that make substantial use of native code, since they pessimistically assume returns from native code can point to arbitrary abstract objects. For demanding, whole-program clients such as static taint analysis, such imprecision generates a huge number of false positives, since every abstract object that leaks into missing code becomes aliased with every return value from missing code. Even with such coarse assumptions, their runtime overhead can be higher than 300\%, which is not suitable for use in production code. In contrast, our analysis is both completely precise and incurs reasonable overhead.

There has also been work identifying bugs~\cite{jin2012bugredux,jin2013f3,liblit2003bug} and information leaks~\cite{austin2012multiple,devriese2010noninterference,enck2014taintdroid} by monitoring production executions. Our work similarly monitors production code to identify unsoundness that can be used to find bugs, information flows, and so forth, but our approach differs in that we aim to use the reported counterexamples to compute \emph{static} points-to sets that are eventually sound; these points-to sets can be used with any client.

\paragraph{Specification inference.}

There has been recent work on inferring specifications, e.g., purely static approaches that interact with a human analyst~\cite{zhu2013automated,bastani2015specification,albarghouthi2016maximal}, and approaches that rely on dynamic traces~\cite{clapp2015modelgen}. Purely static approaches can give certain soundness guarantees, but suffer from imprecision and rely heavily on interaction. In contrast, dynamic approaches are fully automatic, but necessarily incomplete since dynamic analysis is an underapproximation. Our goal is to develop a fully automatic approach where runtime checks are used to detect when specifications are missing.
Furthermore,~\cite{ali2013averroes} enables sound callgraph analysis using only information available in the library interface by using the \emph{separate compilation assumption}, which says that the library can be compiled separately from the program. This assumption is similar to our disjoint fields assumption (with extensions to shared fields and callbacks), in particular, we assume that the only information about the program ``known'' to the library are fields and methods that appear in the library interface. While the callgraph can be computed with reasonable precision using pessimistic assumptions, the same is not true of points-to edges.

\paragraph{Static points-to analysis.}
There is a large literature on static points-to analysis~\cite{shivers1991control,andersen1994program,wilson1995efficient,milanova2002parameterized,whaley2004cloning,sridharan2006refinement}. Our focus is on the new problem of automatic inference of precise points-to information when some of the code is missing.

\paragraph{Static information flow.}
Static information flow analysis has been applied previously to the verification of security policies~\cite{arzt2014flowdroid,fuchs2009scandroid,livshits2005finding,tripp2009taj,xie2006static}. All of these approaches depend on alias analysis, and many use specifications to improve precision and scalability. Our techniques for automatically synthesizing points-to specifications can make implementing any static analysis for large software systems, including information flow analysis, more practical.

\paragraph{Synthesis.}
Program synthesis has also been applied to inferring specifications from dynamic traces~\cite{heule2015mimic}. This approach requires fine-grained instrumentation (specifically, leveraging features of the Javascript language to obtain alias traces), but they recover all method functionality. They accomplish this using MCMC on a restricted space of potential specifications. Our approach requires significantly less instrumentation, but our goal is only to recover aliasing behaviors, and our specifications are furthermore flow insensitive. There have been other approaches to synthesizing programs from traces~\cite{lau2003learning,gulwani2011automating}. See~\cite{heule2015mimic} for a detailed discussion.

\section{Conclusion}

We have described an approach to points-to analysis when code is missing. Our approach is completely precise and fully automatic, and while it forgoes ahead-of-time soundness, it achieves eventual soundness by using runtime checks in production code. We implement our approach in a tool called \toolname to compute points-to sets for Android apps, where the Android framework is missing, and show that our approach achieves low runtime overhead and data usage on almost all apps in a large benchmark suite.

%
%
\bibliographystyle{splncs03}
\bibliography{paper}

\clearpage
\appendix
\section{Soundness and Minimality of Optimized Monitoring}

We prove the propositions in Section~\ref{sec:optimizedmonitoring} that show that our optimized monitoring schemes are sound. First, Proposition~\ref{prop:sufficient} says that we only need to monitor function call returns (in addition to abstract objects). Second, Proposition~\ref{prop:betteralloc} says we only need to monitor abstract objects that leak to missing code (in addition to function call returns).

Finally, we prove Proposition~\ref{prop:minimal}, which describes the sense in which our proposed instrumentation scheme is minimal.

\subsection{Proof of Proposition~\ref{prop:sufficient}}
\label{sec:sufficientproof}

We need to show that if $x\hookrightarrow o$ occurs during execution, then we either report $x\hookrightarrow o$ or $x\hookrightarrow o$ can be statically derived from a reported edge. Suppose $x\hookrightarrow o$ occurs because $x\hookrightarrow\bar{o}$ during execution, where $\bar{o}\leadsto o$. Because we instrument every allocation (and we have assumed that there are no allocations in missing code), we detect $\bar{o}\leadsto o$. Therefore, it suffices to show that our instrumentation reports $x'\hookrightarrow\bar{o}$, where $x\hookrightarrow o$ can be derived from $x'\hookrightarrow o$. We prove by induction on the execution trace:
\begin{itemize}
\item Allocation: If $\bar{o}$ is assigned to $x$ at statement $x\gets X()$, then $\bar{o}\leadsto o=(x\gets X())$, so $x\hookrightarrow o$ cannot be missing (since it is derived by rule 1 in Figure~\ref{fig:pointsto}).
\item Assignment: If $\bar{o}$ is assigned to $x$ at statement $x\gets y$, then at this point in the execution, $y\hookrightarrow\bar{o}$. By induction, we derive $y\hookrightarrow o$ statically, so by rule 2 in Figure~\ref{fig:pointsto}, we statically derive $x\hookrightarrow o$.
\item Load: Suppose that $\bar{o}$ is assigned to $x$ at statement $x\gets y.f$, where $y\hookrightarrow\bar{o}'$ at this point in the execution. Furthermore, at a prior point in the execution, $\bar{o}$ must have been stored into the field $f$ of $\bar{o}'$ via a statement $z.f\gets w$, where $z\hookrightarrow\bar{o}'$ and $w\hookrightarrow\bar{o}$ at this point in the execution. By our disjoint fields assumption, $z.f\gets w$ must be in available code, so $z$ and $w$ are visible. Therefore, by induction, we statically derive $y\hookrightarrow o'$, $z\hookrightarrow o'$, and $w\hookrightarrow o$, where $\bar{o}'\leadsto o'$ (note that we observe $\bar{o}'\leadsto o'$ for the same reason we observe $\bar{o}\hookrightarrow o$). Therefore, by rule 3 in Figure~\ref{fig:pointsto}, we statically derive $x\hookrightarrow o$.
\item Store: A store statement cannot assign $\bar{o}$ to $x$.
\item Function call: If $\bar{o}$ is assigned to $x$ at statement $x\gets m(y)$, then we record $x\hookrightarrow\bar{o}$, since we monitor all such statements.
\end{itemize}
Therefore, $M_{\text{min}}$ is sound.~$\square$

\subsection{Proof of Proposition~\ref{prop:betteralloc}}
\label{sec:betterallocproof}

We now prove Proposition~\ref{prop:betteralloc}. Suppose that a missing points-to edge $x\hookrightarrow o$ occurs during an execution. Without loss of generality, assume that $x\hookrightarrow o$ is the first missing points-to edge to occur in the execution trace. We claim that our instrumentation $\tilde{M}_{\text{min}}$ reports $x\hookrightarrow o$. Our proof of this claim proceeds in two steps:
\begin{itemize}
\item First, suppose that $x\hookrightarrow o$ occurs because $x\hookrightarrow\bar{o}$, where $\bar{o}\leadsto o$. Then, we show that $\bar{o}$ must have been assigned to $x$ via a call to a missing function $x\gets m(y)$.
\item Second, we show that for any $\bar{o}$ assigned to $x$ via a call to a missing function $x\gets m(y)$, we must have $y'\hookrightarrow\bar{o}$, where $y'$ is an argument passed to missing code via a call $x'\gets m'(y')$ to a missing function $m'$.
\end{itemize}
Intuitively, the first claim says that a missing points-to edge $x\hookrightarrow\bar{o}$ can only ``introduced'' into the program as a result of a call to a missing function, and the second claim says that a concrete object $\bar{o}$ can only be returned by a call to a missing function if it was previously passed as the argument to a (possibly different) missing function.

We show the first claim; i.e., that $\bar{o}$ must have been assigned to $x$ via a call $x\gets m(y)$ to a missing function $m\in\mathcal{M}$. We proceed by induction on the execution trace:
\begin{itemize}
\item Allocation: If $x\gets X()$ assigns $\bar{o}$ to $x$, then $\bar{o}\leadsto o=(x\gets X())$, so $x\hookrightarrow o$ cannot be missing.
\item Assignment: If $x\gets y$ assigns $\bar{o}$ to $x$, then $y\hookrightarrow\bar{o}$, so $y\hookrightarrow o$ must also have been missing (or we would have derived $x\hookrightarrow o$ statically).
\item Load: Suppose $x\gets y.f$ assigns $\bar{o}$ to $x$. Then, assuming $y\hookrightarrow\bar{o}'$ at this point in the execution, $\bar{o}$ must have been stored into field $f$ of $\bar{o}'$ by a statement $z.f\gets w$ (which is also in available code by our disjoint fields assumption), where $z\hookrightarrow\bar{o}'$ and $w\hookrightarrow\bar{o}$. One of the edges $y\hookrightarrow o'$, $z\hookrightarrow o'$, and $w\hookrightarrow o$ must have been missing (or we would have derived $x\hookrightarrow o$ statically).
\item Store: Such a statement cannot assign $\bar{o}$ to $x$.
\end{itemize}

Now, we show the second claim; i.e., that $\bar{o}$ must have been assigned to some argument $y'$ passed to missing code via a function call $x'\gets m'(y')$. Our proof uses the points-to set $\hat{\Pi}$ computed by including all missing code in the static analysis in Figure~\ref{fig:pointsto} (with $\Pi_{\text{miss}}=\emptyset$). We cannot compute $\hat{\Pi}$ since we do not have the missing code, but $\hat{\Pi}$ is sound, so in particular, $x\hookrightarrow o\in\hat{\Pi}$ reported by our instrumentation, we have $x\hookrightarrow o\in\hat{\Pi}$.

We prove the following stronger claim. Suppose that the points-to edge $x'\hookrightarrow\bar{o}$ occurs during execution, where $x'$ is either in missing code or equal to the parameter $p_m$ or return value $r_m$ of a missing function $m$, and $\bar{o}$ is allocated at allocation statement $o$ in available code. Then, at a prior point in the execution, $y'\hookrightarrow\bar{o}$ for some argument $y'$ (in visible code) passed to missing code via a call $x'\gets m'(y')$ to missing function $m'$. We prove by induction on the execution trace:
\begin{itemize}
\item Allocation: Note that $\bar{o}$ cannot be assigned to $x'$ via an allocation statement $x'\gets X()$, since we have assumed that $\bar{o}$ is allocated in available code.
\item Assignment: If $x'\gets y'$ assigns $\bar{o}$ to $x'$, then either $y'$ is also missing, in which case our claim follows by induction, or $x'=p_{m'}$ is a parameter and the (visible) variable $y'$ is an argument passed to missing code via a call $x''\gets m'(y')$ to missing function $m'$, so again our claim follows.
\item Load: Suppose that $x'\gets y'.f$ assigns $\bar{o}$ is assigned to $x'$. Furthermore, suppose that $y'\hookrightarrow\bar{o}'$ at this point in the execution; then, $\bar{o}$ must have been stored into field $f$ of $\bar{o}'$ by statement $z.f\gets w$ at a prior point in the execution, where $z\hookrightarrow\bar{o}'$ and $w\hookrightarrow\bar{o}$. By our separation of fields assumption, $z.f\gets w$ must be in missing code. Since $w$ is in missing code, our claim follows by induction.
\item Store: Such a statement cannot assign $\bar{o}$ to $x$.
\end{itemize}

By our first claim, the first missing points-to edge that occurs during execution is $x\hookrightarrow\bar{o}$ (where $\bar{o}\leadsto o$ is allocated in available code), where $\bar{o}$ is assigned to $x$ via a call $x\gets m(y)$ via a missing function $m$. Letting $x'=r_m$ be the return value $m$, we can apply our second claim, which says that $y'\hookrightarrow\bar{o}$ for some visible variable $y'$ passed to missing code via a call $x'\gets m'(y')$ to a missing function $m'$. By our assumption that $x\hookrightarrow\bar{o}$ is the first missing points-to edge, the points-to edge $y\hookrightarrow\bar{o}$ cannot be missing. In other words, we compute $y\hookrightarrow o\in\Pi$ using our static analysis. Therefore, $o\in\tilde{M}_{\text{alloc}}$ is monitored, from which our result follows.~$\square$

\subsection{Proof of Proposition~\ref{prop:minimal}}
\label{sec:minimalproof}

Our proof depends on the extension to handling allocations in library code described in Section~\ref{sec:proxy}. Assume $\Pi_{\text{miss}}=\emptyset$. Consider any library function return value $r_m$. The library function $m$ could allocate a fresh abstract object to its return value, which would be expressed by the proxy object $p=\{r_m\}\in\mathcal{P}$. In this case, the points-to edge $r_m\hookrightarrow p$ is missing, since it occurs during an execution but is not computed statically. Therefore, if $r_m$ is not monitored, then we would not detect the missing points-to edge $r_m\hookrightarrow p$.

Next, consider any allocation $o$ such that a concrete object $\bar{o}$ is passed as an argument to a library function $m$, i.e., $p_m\hookrightarrow o$. The library function $m$ could assign assign its parameter to its return value. In this case, the points-to edge $r_m\hookrightarrow o$ is missing, since it occurs during an execution but is not computed statically. Furthermore, all other library functions could be no-ops, in which case no other points-to edge is missing. If $o$ is not monitored (but $r_m$ is monitored), then we would not observe $o$ in the program, so we would incorrectly conclude that $r_m$ points to an object allocated in library code. Therefore, $\tilde{M}_{\text{alloc}}$ is minimal as claimed.~$\square$

\section{Guarantees for Proxy Objects}

We prove the propositions in Section~\ref{sec:proxy} that show that the points-to sets we compute with proxy objects are sound and precise. Proposition~\ref{prop:idealsoundness} says that proxy objects are sound, Proposition~\ref{prop:idealprecision} says that ideal proxy objects are precise, and Proposition~\ref{prop:functionfootprint} says that from the perspective of the static analysis, our proxy objects are sound and as precise as ideal proxy objects.

\subsection{Soundness for Ideal Proxy Objects}
\label{sec:idealsoundnessproof}

First, we prove our soundness result for ideal proxy objects. Suppose that $x\hookrightarrow\bar{o}$ during an execution. We prove that $x\hookrightarrow o\in\tilde{\Pi}^*$ (if $\bar{o}$ is allocated at $o$, in which case $\bar{o}\leadsto o$) and $x\hookrightarrow\tilde{p}\in\tilde{\Pi}^*$ (if $\bar{o}$ is allocated in missing code and $\tilde{\phi}(\bar{o})=\tilde{p}$, in which case $\bar{o}\leadsto\tilde{p}$). We prove by induction on the execution trace:
\begin{itemize}
\item Allocation: If $x\gets X()$ assigns $\bar{o}$ to $x$, then $\bar{o}$ is allocated at $o=(x\gets X())$, so we derive $x\hookrightarrow o$ using rule 1 in Figure~\ref{fig:pointsto}.
\item Assignment: If $x\gets y$ assigns $\bar{o}$ to $x$, then by induction, either $y\hookrightarrow o\in\tilde{\Pi}^*$ (if $\bar{o}\leadsto o$) or $y\hookrightarrow\tilde{p}\in\tilde{\Pi}^*$ (if $\bar{o}\leadsto\tilde{p}$). We derive $x\hookrightarrow o$ (in the former case) or $x\hookrightarrow\tilde{p}$ (in the latter case) by rule 2 in Figure~\ref{fig:pointsto}.
\item Load: Suppose $x\gets y.f$ assigns $\bar{o}$ to $x$, and $y\hookrightarrow\bar{o}'$ at this point. Then, $\bar{o}$ must have been stored in field $f$ of $\bar{o}'$ by some statement $z.f\gets w$, where $z\hookrightarrow\bar{o}'$ and $w\hookrightarrow\bar{o}$. By our disjoint fields assumption, $z.f\gets w$ is in available code, so by induction, either $w\hookrightarrow o\in\tilde{\Pi}^*$ (if $\bar{o}\leadsto o$) or $w\hookrightarrow\tilde{p}\in\tilde{\Pi}^*$ (of $\bar{o}\leadsto\tilde{p}$). Furthermore, by induction, either $y\hookrightarrow o'\in\tilde{\Pi}^*$ and $z\hookrightarrow o'\in\tilde{\Pi}^*$ (if $\bar{o}'\leadsto o'\in\mathcal{O}_P$) or $y\hookrightarrow\tilde{p}'\in\tilde{\Pi}^*$ and $z\hookrightarrow\tilde{p}'\in\tilde{\Pi}^*$ (if $\bar{o}'\leadsto\tilde{p}'$). Therefore, we derive $x\hookrightarrow o\in\tilde{\Pi}^*$ (if $\bar{o}\leadsto o$) or $x\hookrightarrow\tilde{p}\in\tilde{\Pi}^*$ (if $\bar{o}\leadsto\tilde{p}$) using rule 3 in Figure~\ref{fig:pointsto}.
\item Store: Note that $\bar{o}$ cannot be assigned to $x$ using this statement.
\item Function call: Suppose $x\gets m(y)$ assigns $\bar{o}$ to $x$. If $\bar{o}\leadsto o$, then $x\hookrightarrow o\in\tilde{\Pi}_{\text{miss}}^*$ is missing, so $x\hookrightarrow o\in\tilde{\Pi}^*$ by rule 4. If $\bar{o}\leadsto\tilde{p}$, then $x\hookrightarrow\tilde{\phi}(\bar{o})\in\tilde{\Pi}^*$ is missing, where $\tilde{p}=\tilde{\phi}(\bar{o})$. Again, $x\hookrightarrow\tilde{p}\in\tilde{\Pi}^*$ by rule 4.
\end{itemize}
The result follows.~$\square$

\subsection{Precision for Ideal Proxy Objects}
\label{sec:idealprecisionproof}

Now, we prove our precision result for ideal proxy objects. Recall that our goal is to prove that (i) if $x\hookrightarrow o\in\tilde{\Pi}^*$, where $o\in\mathcal{O}_P$, then also $x\hookrightarrow o\in\Pi$, and (ii) if $x\hookrightarrow\tilde{p}\in\tilde{\Pi}^*$, where $\tilde{p}=\tilde{\phi}(\bar{o})\in\tilde{\mathcal{P}}$, then $x\hookrightarrow o\in\Pi$, where $o$ is the (missing) allocation statement where $\bar{o}$ was allocated. We proceed by structural induction, first for $x\hookrightarrow o\in\tilde{\Pi}^*$:
\begin{itemize}
\item Allocation: If $x\hookrightarrow o$ is derived using rule 1 in Figure~\ref{fig:pointsto}, then $o=(x\gets X())$, so $x\hookrightarrow o\in\Pi$ by rule 1 as well.
\item Assignment: If $x\hookrightarrow o$ is derived using rule 2 in Figure~\ref{fig:pointsto}, then $x\gets y$ and $y\hookrightarrow o\in\tilde{\Pi}^*$. By induction, $y\hookrightarrow o\in\Pi$, so $x\hookrightarrow o\in\Pi$ is derived using rule 2 in Figure~\ref{fig:pointsto}.
\item Load/store: Suppose that $x\hookrightarrow o$ is derived using rule 3 in Figure~\ref{fig:pointsto}---i.e., from premise
\begin{align*}
x\gets y.f,\sss z.f\gets w,\sss w\hookrightarrow o,\sss y\hookrightarrow o',\sss z\hookrightarrow o',
\end{align*}
where $o'\in\mathcal{O}_P$, or premise
\begin{align*}
x\gets y.f,\sss z.f\gets w,\sss w\hookrightarrow o,\sss y\hookrightarrow\tilde{p}',\sss z\hookrightarrow\tilde{p}',
\end{align*}
where $\tilde{p}'\in\tilde{\mathcal{P}}$. In either case, by induction, $w\hookrightarrow o\in\Pi$. In the former case, $y\hookrightarrow o'\in\Pi$ and $z\hookrightarrow o'\in\Pi$ by induction, so we derive $x\hookrightarrow o\in\Pi$ using rule 3 in Figure~\ref{fig:pointsto}. In the latter case, suppose that $\tilde{p}'=\tilde{\phi}(\bar{o}')$, and let $o'$ be the statement at which $\bar{o}'$ was allocated. By induction, $y\hookrightarrow o'\in\Pi$ and $z\hookrightarrow o'\in\Pi$, so we again derive $x\hookrightarrow o\in\Pi$ using rule 3 in Figure~\ref{fig:pointsto}.
\item Missing: If $x\hookrightarrow o\in\tilde{\Pi}_{\text{miss}}$, then $x\hookrightarrow\bar{o}$ during a concrete execution, where $\bar{o}$ was allocated at $o$. Therefore, any sound points-to analysis must compute $x\hookrightarrow o$.
\end{itemize}
Second, we prove the claim for $x\hookrightarrow\tilde{p}\in\tilde{\Pi}^*$:
\begin{itemize}
\item Allocation: We cannot derive $x\hookrightarrow\tilde{p}$ using rule 1 in Figure~\ref{fig:pointsto}.
\item Assignment: If $x\hookrightarrow\tilde{p}$ is derived using rule 2 in Figure~\ref{fig:pointsto}, then $x\gets y$ and $y\hookrightarrow\tilde{p}\in\tilde{\Pi}^*$. Suppose that $\tilde{p}=\tilde{\phi}(\bar{o})$, and $\bar{o}$ is allocated at statement $o$. By induction, $y\hookrightarrow o\in\Pi$, so we derive $x\hookrightarrow o\in\Pi$ using rule 2 in Figure~\ref{fig:pointsto}.
\item Load/store: Suppose that $x\hookrightarrow\tilde{p}$ is derived using rule 3 in Figure~\ref{fig:pointsto}---i.e., from premise
\begin{align*}
x\gets y.f,\sss z.f\gets w,\sss w\hookrightarrow\tilde{p},\sss y\hookrightarrow o',\sss z\hookrightarrow o',
\end{align*}
where $o'\in\mathcal{O}_P$, or premise
\begin{align*}
x\gets y.f,\sss z.f\gets w,\sss w\hookrightarrow\tilde{p},\sss y\hookrightarrow\tilde{p}',\sss z\hookrightarrow\tilde{p}',
\end{align*}
where $\tilde{p}'\in\tilde{\mathcal{P}}$. In either case, by induction, $w\hookrightarrow o\in\Pi$, where $\tilde{p}=\tilde{\phi}(\bar{o})$ and $o$ is the statement at which $\bar{o}$ is allocated. In the former case, by induction, $y\hookrightarrow o'\in\Pi$ and $z\hookrightarrow o'\in\Pi$, so we derive $x\hookrightarrow o\in\Pi$ using rule 3 in Figure~\ref{fig:pointsto}. In the latter case, let $o'$ be the allocation statement of $\bar{o}'$, where $\tilde{p}'=\tilde{\phi}(\bar{o}')$. By induction, $y\hookrightarrow o'\in\Pi$ and $z\hookrightarrow o'\in\Pi$, so we again derive $x\hookrightarrow o\in\Pi$ using rule 3 in Figure~\ref{fig:pointsto}.
\end{itemize}
The result follows.~$\square$

\subsection{Soundness and Precision for Proxy Objects}
\label{sec:functionfootprintproof}

In this section, we prove Proposition~\ref{prop:functionfootprint}, which essentially says that our proxy object $\phi$ mapping is as precise as the ideal proxy object mapping $\tilde{\phi}$. Suppose that $\tilde{p}=\tilde{\phi}(\bar{o})$ and $p=\phi(\bar{o})$ for concrete object $\bar{o}$. We prove that $x\hookrightarrow\tilde{p}\in\tilde{\Pi}^*$ if and only if $x\hookrightarrow p\in\Pi^*$.

First, we prove that if $x\hookrightarrow\tilde{p}\in\tilde{\Pi}^*$ (resp., $x\hookrightarrow o\in\tilde{\Pi}^*$), then $x\hookrightarrow p\in\Pi^*$ (resp., $x\hookrightarrow o\in\Pi^*$). We proceed by structural induction on the derivation of $x\hookrightarrow\tilde{p}$ (resp., $x\hookrightarrow o$) in $\tilde{\Pi}^*$:
\begin{itemize}
\item Allocation: Note that $x\hookrightarrow\tilde{p}$ cannot be derived using rule 1 in Figure~\ref{fig:pointsto} since it only applies to allocations $o\in\mathcal{O}_P$. For $x\hookrightarrow o$, we must have $o=(x\gets X())$, in which case we derive $x\hookrightarrow o\in\Pi^*$ as well using rule 1.
\item Assignment: If $x\hookrightarrow\tilde{p}$ (resp., $x\hookrightarrow o$) is derived from $x\gets y$, then $y\hookrightarrow\tilde{p}\in\tilde{\Pi}^*$ (resp., $y\hookrightarrow o\in\tilde{\Pi}^*$), so by induction $y\hookrightarrow p\in\Pi^*$ (resp., $y\hookrightarrow o\in\Pi^*$). Therefore, we derive $x\hookrightarrow p\in\Pi^*$ (resp., $x\hookrightarrow o\in\Pi^*$) using rule 2 in Figure~\ref{fig:pointsto}.
\item Load/store: Suppose $x\hookrightarrow\tilde{p}$ is derived using rule 3 in Figure~\ref{fig:pointsto} from premise
\begin{align*}
x\gets y.f,~z.f\gets w,~y\hookrightarrow\tilde{p}'\in\tilde{\Pi}^*,~z\hookrightarrow\tilde{p}'\in\tilde{\Pi}^*,~w\hookrightarrow\tilde{p},
\end{align*}
where $\tilde{p}'=\tilde{\phi}(\bar{o}')\in\tilde{\mathcal{P}}$. Let $p'=\phi(\bar{o}')$. Then, by induction, we derive $y\hookrightarrow p'\in\Pi^*$, $z\hookrightarrow p'\in\Pi^*$, and $w\hookrightarrow p\in\Pi^*$, so we derive $x\hookrightarrow p\in\Pi^*$ using rule 3 in Figure~\ref{fig:pointsto}. The cases where $\tilde{p}$ is instead $o\in\mathcal{O}_P$ and/or $\tilde{p}'$ is instead $o'\in\mathcal{O}_P$ follow similarly.
\item Missing: Suppose $x\hookrightarrow\tilde{p}$ is derived using rule 4 in Figure~\ref{fig:pointsto}, so $x\hookrightarrow\tilde{p}\in\tilde{\Pi}_{\text{miss}}^*$. Then, $x$ is in the dynamic footprint of $\bar{o}$ (i.e., $x\hookrightarrow\bar{o}$ during execution). We show below that if $x$ is in the dynamic footprint of $\bar{o}$, then we derive $x\hookrightarrow p\in\Pi^*$. Note that $x\hookrightarrow o$ cannot be derived using this rule.
\end{itemize}
We prove the claim in the last case by induction on the execution trace:
\begin{itemize}
\item Allocation: Note that $\bar{o}$ cannot be assigned to $x$ using an allocation $x\hookrightarrow X()$, since we have assumed that $\bar{o}$ is allocated in missing code.
\item Assignment: If $x\gets y$ assigns $\bar{o}$ is to $x$, then $y\hookrightarrow\bar{o}$ at that point in the execution. By induction, $y\hookrightarrow p\in\Pi^*$, so we derive $x\hookrightarrow p\in\Pi^*$ using rule 2 in Figure~\ref{fig:pointsto}.
\item Load: Suppose that $x\gets y.f$ assigns $\bar{o}$ to $x$, and $y\hookrightarrow\bar{o}'$ at that point in the execution. Then, at a previous point in the execution, $\bar{o}$ must have been assigned to field $f$ of $\bar{o}'$ by a statement $z.f\gets w$, where $z\hookrightarrow\bar{o}'$ and $w\hookrightarrow\bar{o}$. By induction on the execution trace, we have $w\hookrightarrow p\in\Pi^*$.

If $\bar{o}'$ is allocated in missing code, then by Proposition~\ref{prop:idealsoundness}, $y\hookrightarrow\tilde{p}'\in\tilde{\Pi}^*$ and $z\hookrightarrow\tilde{p}'\in\tilde{\Pi}^*$, where $\tilde{p}'=\tilde{\phi}(\bar{o}')$. Then, by structural induction (on the derivation of $x\hookrightarrow\tilde{p}$), we have $y\hookrightarrow p'\in\Pi^*$ and $z\hookrightarrow p'\in\Pi^*$, where $p'=\phi(\bar{o}')$. Otherwise, if $\bar{o}'$ is allocated at $o'\in\mathcal{O}_P$, then by Proposition~\ref{prop:idealsoundness}, $y\hookrightarrow o'\in\tilde{\Pi}^*$ and $z\hookrightarrow o'\in\tilde{\Pi}^*$. Then, by structural induction (on the derivation of $x\hookrightarrow\tilde{p}$), we have $y\hookrightarrow o'\in\Pi^*$ and $z\hookrightarrow o'\in\Pi^*$. Either way, we derive $x\hookrightarrow p\in\Pi^*$ using rule 3 in Figure~\ref{fig:pointsto}.
\item Store: Note that $\bar{o}$ cannot be assigned to $x$ using this statement.
\item Function call: If $x\gets m(y)$ assigns $\bar{o}$ to $x$, then $m$ is in the function footprint of $\bar{o}$, so $x\hookrightarrow p\in\Pi_{\text{miss}}^*\subseteq\Pi^*$.
\end{itemize}
Therefore, we have shown the forward implication. Next, we show if $x\hookrightarrow p\in\Pi^*$ (resp., $x\hookrightarrow o\in\Pi^*$), then $x\hookrightarrow\tilde{p}\in\tilde{\Pi}^*$ (resp., $x\hookrightarrow o\in\Pi^*$). We prove by structural induction on the derivation of $x\hookrightarrow p$:
\begin{itemize}
\item Allocation: As before, $x\hookrightarrow p$ cannot be derived using rule 1 in Figure~\ref{fig:pointsto}, and for $x\hookrightarrow o$, we must have $o=(x\gets X())$, in which case we derive $x\hookrightarrow o\in\tilde{\Pi}^*$ as well using rule 1.
\item Assignment: If $x\hookrightarrow p$ (resp., $x\hookrightarrow o$) is derived from $x\gets y$, then $y\hookrightarrow p\in\Pi^*$ (resp., $y\hookrightarrow o\in\Pi^*$), so by induction, $y\hookrightarrow\tilde{p}\in\tilde{\Pi}^*$ (resp., $y\hookrightarrow o\in\tilde{\Pi}^*$), in which case we derive $x\hookrightarrow\tilde{p}\in\tilde{\Pi}^*$ using rule 1 in Figure~\ref{fig:pointsto}.
\item Load/store: Suppose $x\hookrightarrow p$ (resp., $x\hookrightarrow o$) is derived from premise
\begin{align*}
\begin{array}{c}
x\gets y.f,\sss z.f\gets w,\sss y\hookrightarrow p'\in\Pi^*,\\
z\hookrightarrow p'\in\Pi^*,\sss w\hookrightarrow p\in\Pi^*,
\end{array}
\end{align*}
where $p'=\phi(\bar{o}')$. Let $\tilde{p}'=\tilde{\phi}(\bar{o}')$. By induction, $y\hookrightarrow\tilde{p}'\in\tilde{\Pi}^*$, $z\hookrightarrow\tilde{p}'\in\tilde{\Pi}^*$, and $w\hookrightarrow\tilde{p}\in\tilde{\Pi}^*$, so we derive $x\hookrightarrow\tilde{p}\in\tilde{\Pi}^*$ using rule 3 in Figure~\ref{fig:pointsto}.
\item Missing: Suppose $x\hookrightarrow p\in\Pi_{\text{miss}}^*$, so $x\gets m(y)$, where $m$ is in the function footprint of $\bar{o}$. Then, $x$ is in the dynamic footprint of $\bar{o}$, so $x\hookrightarrow\tilde{p}\in\tilde{\Pi}_{\text{miss}}^*$, so we derive $x\hookrightarrow\tilde{p}\in\tilde{\Pi}^*$ using rule 4 in Figure~\ref{fig:pointsto}.
\end{itemize}
Therefore, the backwards implication follows, so we are done.~$\square$

\end{document}